\documentclass[a4paper,10pt]{article}

\usepackage{rotating}

\usepackage{color}
\usepackage{graphicx}
\usepackage{amsmath}
\usepackage{graphics}
\usepackage{amsthm}
\usepackage{amsfonts}
\usepackage{url}
\usepackage{authblk}
\usepackage{longtable}
\usepackage[margin=0.75in]{geometry}

\def\blfootnote{\xdef\@thefnmark{}\@footnotetext}
\long\def\symbolfootnote[#1]#2{\begingroup%
\def\thefootnote{\fnsymbol{footnote}}\footnote[#1]{#2}\endgroup}

\newtheorem{definition}{Definition}

\newcommand{\LV}{Low Voltage }
\newcommand{\MV}{Medium Voltage }
\newcommand{\HV}{High Voltage }
\newcommand{\CPL}{Characteristic Path Length }

\newcommand{\PG}{Power Grid }
\newcommand{\PGs}{Power Grids }
\newcommand{\etal}{\textit{et al.}}
\newcommand{\G}{Grid }
\newcommand{\Gs}{Grids }
\newcommand{\sw}{small-world }
\newcommand{\MLV}{Medium and Low Voltage }
\newcommand{\CNA}{Complex Network Analysis }
\newcommand{\Exp}{Exponential }
\newcommand{\Pl}{Power-law }

\begin{document}
\author{Giuliano~Andrea~Pagani} 
\author{Marco~Aiello}

\title{The \PG as a Complex Network: a Survey}

\affil{Distributed Systems Group\\Johann Bernoulli Institute for Mathematics and Computer Science
\\University of Groningen\\ Groningen, The Netherlands\\
\vspace{0.3cm}
email: \url{{g.a.pagani,m.aiello}@rug.nl}\\
 \url{http://www.cs.rug.nl/ds/}
}

\maketitle

\begin{abstract}
The statistical tools of Complex Network Analysis are of great use to
understand salient properties of complex systems, may these be natural
or pertaining human engineered infrastructures. One of these that is
receiving growing attention for its societal relevance is that of electricity distribution. In this paper, we present a survey of the
most important scientific studies investigating the properties of
several \PGs infrastructures using \CNA techniques and
methodologies. 
We categorize and explore the most relevant
literature works considering general topological
properties, differences between the various graph-related indicators
and reliability aspects. 
\end{abstract}

\section{Introduction}\label{sec:intro}
{\em \CNA (CNA)} is a relatively young field of research. The first
systematic studies appeared in the late 1990s~\cite{Watts98,
  Strogatz2001, Barabasi1999,Albert2000} having the goal of studying the
properties of large networks that behave as complex systems. The
research owes a great deal of its foundations to the seminal work on
Random Graphs of Erd{\H{o}}s and
R{\'{e}}nyi~\cite{Erdos1959,Erdos1960} who studied asymptotic
properties of stochastic graph processes.
 \CNA has been used in many different fields
of knowledge, from biology~\cite{Jeong2000} to
chemistry~\cite{Doye2002}, from linguistics to social
sciences~\cite{Milgram69}, from telephone call patterns~\cite{Aiello2000} to
computer networks~\cite{Faloutsos1999} and
web~\cite{albert99,Donato2007} to virus
spreading~\cite{kephart91,colizza07,Gautreau2008} to
logistics~\cite{Latora2002,Guimer2004,Colizza2006} and also
inter-banking systems~\cite{Nationalbank2003}. Men-made infrastructures
are especially interesting to study under the \CNA lenses, especially
when they are large scale and grow in a decentralized and independent
fashion, thus not the result of a global, but rather of many local
autonomous designs. The  \PG is a prominent example. But what do we
mean by \PG in the context of the present treatment?

We focus on the electricity transmission and distribution \PG as it is essential for
today's society as an enabling infrastructure, but also its efficiency
and working has major consequences, among other things, for the environment. Blackouts
seem to have a special role in reminding us of the importance of the
Grid and how much we give its availability for granted. From the
technological point of view, the electrical system and
\PG involve many scientific knowledge areas that contribute to the
design, operations
and analysis of power systems: Physics (electromagnetism, classical
mechanics), Electrical engineering (AC circuits and phasors, 3-phase
networks, electrical systems control theory) and Mathematics (linear
algebra, differential equations). Traditional studies 
tend to have a ``local'' view of the Grid, e.g., defining how to
design a transformer and predicting its functioning. Typically, studies
tend to focus on the physical and electrical properties
(e.g.,~\cite{Anghel06}), or the characteristics of the \PG as a
complex dynamical system~\cite{Dobson012}, or again, the control
theory aspects~\cite{Hiskens97}. The move from a ``local'' to a
``global'' view of the \PG as a complex system is possible by
resorting to \CNA and statistical graph theory.

The goal of the present treatment is to provide a survey and compare the most
well-known scientific studies conducted using \CNA techniques
concerning \PG systems. We consider several parameters to assess the
differences between the various studies and try to enucleate the most
important aspects of each study. We start by introducing the
methods and metrics that are evaluated in this work (Section~\ref{sec:method}); the section
contains the basic definitions and simple examples in order to
establish a common background.
Section~\ref{sec:relwks} provides the main characteristics of all the
studies. The actual comparison of these using CNA metrics are reported
and discussed in Section~\ref
{sec:discConc} which also provides a conclusion to the paper.

\section{Background and Survey Methodology}\label{sec:method}

\begin{figure}[htbp]
\centering
\framebox[0.5\linewidth]
{
\begin{minipage}{0.5\linewidth} 
  \centering
\includegraphics[width=0.3\textwidth]{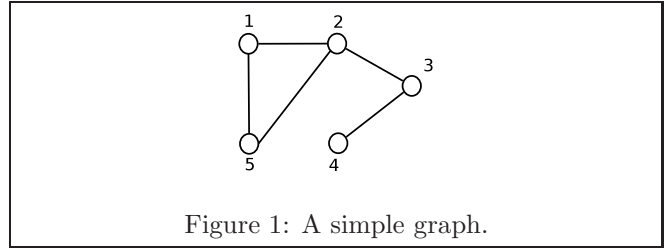}
 \caption{A simple graph.}
 \label{fig:graphEx}
 \end{minipage}
 }
\end{figure}

Before going further analyzing the various studies in detail, some common definitions need to be stated in order to have a common ground. The essential concepts from the graph theory are also given to provide a common basis. For each graph property described, a concrete example on a small graph, as shown in Figure~\ref{fig:graphEx}, is provided to better understand property's application.

As described in Section~\ref{sec:intro}, all the works that are examined in the present manuscript consider the \PG networks as graphs following the mathematical meaning of the term. 

\begin{definition}[Graph]\label{def:graph}
A {\em graph} $G$ is a pair of sets G(V,E) where $V$ is
the set of vertexes and $E$ is the set edges. An edge $e_{i,j}$ is a
pair of vertexes $(v_i,v_j)$. If $(v_i,v_j)\in E$ then $v_i\textrm{
  and }v_j$ are said to be {\em adjacent} or {\em neighboring} and are
called {\em end-vertexes} of the edge.
\end{definition}

\noindent
Considering the Power Grid, the sets composing the graph assume particular interest from an operational and physical point of view. The physical components of the \PG assume a meaning in the theoretical representation of the \PG as a graph according the following interpretation.
\begin{definition}[Power Grid graph]\label{def:PGgraph}
A {\em Power Grid graph} is a graph $G(V,E)$ such that each element 
$v_i \in V$ is either a substation, transformer, or consuming unit of a
physical Power Grid. There is an edge $e_{i,j}=(v_i,v_j) \in E$
between two nodes if there is a physical cable connecting directly the elements represented by $v_i$ and $v_j$.
\end{definition}
Therefore, following the properties of the graphs is an interesting first way of categorizing the \PG under analysis. In particular, distinctions can be made regarding the \emph{order} and \emph{size} of the graph. \textit{Order} is the number of vertexes composing the graph, while \textit{size} is the number of edges in the same graph. More formally:
\begin{definition}[Order and size of a  graph]
Given the graph $G$ the order is given by $N=|V|$, while the
size is given by $M=|E|$.
\end{definition}

\vspace*{1cm}
\begin{center}
\centering
\framebox[0.75\linewidth]{
\begin{minipage}{0.70\linewidth}
\textbf{Example:}

The graph $G$ shown in Figure~\ref{fig:graphEx} is characterized by the set of vertexes $V$:
\[
V=\{1,2,3,4,5\}
\]
and by the set of edges $E$:
\[
E=\{(1,2),(1,5),(2,3),(2,5),(3,4)\}
\]
the order and size of $G$ are $|V|=5$ and $|E|=5$ respectively.\\
\end{minipage}
}
\end{center}
\vspace*{1cm}
An important property that characterizes a vertex of a graph is the degree, that is the number of vertexes the node is adjacent to. More formally, this quantity is defined as:
\begin{definition}[Neighborhood and degree]
The set of vertexes adjacent to a vertex $v \in V$ represents the \emph{neighborhood} of $v$ that is denoted by $\Gamma_v$.
The \emph{degree} of $v$ is $d(v)=|\Gamma_v|$.
\end{definition}
From \textit{order} and \textit{size} it is possible to have a global value for the connectivity of the vertexes of the graph, known
as {\em average node degree} . That is $<\!k\!>=\frac{2M}{N}$.
\vspace*{1cm}
\begin{center}
\framebox[0.75\linewidth]{
\begin{minipage}{0.70\linewidth}
\textbf{Example:}

The neighborhood of vertex 1 in Figure~\ref{fig:graphEx} is:
\[
 \Gamma(1)=\{2,5\}
\]
its degree is:
\[
 d(1)=|\Gamma(1)|=2
\]
The average node degree for $G$ is:
\[
<\!k\!>=\frac{2\cdot 5}{5}=2
\]
\end{minipage}
}
\end{center}
\vspace*{1cm}
Usually, it is not essential to have the specific information regarding the node degree of only a certain node or the average degree. It is more interesting understand the the overall characteristics of a graph considering its statistical measures. In particular, one measure is the node degree probability distribution.
More formally,
\begin{definition}[Node degree distribution]\label{def:ndd}
Consider the degree $k$ of a node in a graph as a random variable, the
function 
\[
N_k=\{v\in G:\: d(v)=k\}
\]
is called {\em probability node degree distribution}.
\end{definition}

\noindent
The shape of the distribution is a salient characteristic of the
network. For the Power Grid,
the shape is typically either exponential or a Power-law.
More precisely, an exponential node degree ($k$) distribution has a
fast decay in the probability of having nodes with relative high node
degree. It follows the relation:
\[
 P(k)=\alpha e^{\beta k}
\]
where $\alpha$ and $\beta$ are parameters of the specific network
considered. While a \Pl distribution has a slower decay with higher probability of having nodes with high node degree. It is expressed by the relation:
\[
 P(k)=\alpha k^{-\gamma}
\]
where $\alpha$ and $\gamma$ are parameters of the specific network
considered. We remark that the graphs considered in the \PG domain are
usually large, although finite, in terms of order and size thus
providing limited and finite probability distributions.

%
%

%
%

%
The node degree distribution gives some information about the static situation of the network, but it does not give any information about the paths that can be followed in the graph to move from one node to another. To investigate properties dealing with paths between nodes, that for a \PG graph are important to assess which nodes have to sustain the highest flow of energy, some further concepts are essential.

The concepts of \emph{path} and \emph{path length} are crucial to understand the way two vertexes are connected.
\begin{definition}[Path and  path length]
A \emph{path} of G is a subgraph $P$ of the form:
\[V(P)=\{x_0,x_1,\ldots,x_l\}, \hspace{10mm}
  E(P)=\{(x_0,x_1),(x_1,x_2),\ldots,(x_{l-1},x_l)\}.
\]
such that $V(P)\subseteq V\textrm{ and }E(P)\subseteq E$. The vertexes
$x_0$ and $x_l$ are \emph{end-vertexes} of $P$ and $l=|E(P)|$ is the
\emph{length} of $P$. A graph is {\em connected} if for any two
distinct vertexes $v_i,v_j\in V$ there is a finite path from $v_i$ to $v_j$.\\
\end{definition}

\begin{definition}[Distance]
Given a graph $G$ and vertexes $v_i$ and $v_j$, their \emph{distance} $d(v_i,v_j)$ is
the minimal length of any $v_i-v_j$ path in the graph. If there is no
$v_i-v_j$ path then it is conventionally  set to $d(v_i,v_j)=\infty$.\\
\end{definition}
\begin{definition}[Shortest path]
Given a graph $G$ and vertexes $v_i$ and $v_j$ the {\em shortest path} is
the the path corresponding to the minimum of to the set $\{|P_1|,|P_2|,\ldots,|P_k|\}$ containing the
lengths of all paths for which $v_i$ and $v_j$ are the end-vertexes.
\end{definition}

\begin{figure}[htbp]
 \centering
\framebox[0.5\linewidth]
{
\begin{minipage}{0.5\linewidth} 
  \centering
\includegraphics[width=0.45\textwidth]{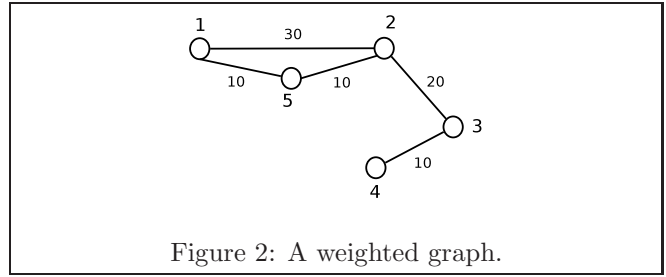}
 \caption{A weighted graph.}
 \label{fig:weightedGraph}
 \end{minipage}
 }
\end{figure}

\vspace*{1cm}
\begin{center}
\framebox[0.75\linewidth]
{
\begin{minipage}{0.70\linewidth}
\textbf{Example:}

Example of paths between vertex 1 and vertex 4 are:
\[
P_{1,4}=\{1-2-3-4\}
\]
whose length $l_{P}=3$ but also:
\[
P^{'}_{1,4}=\{1-5-2-3-4\}
\]
is a valid path whose length $l_{P^{'}}=4$, therefore the shortest path between vertex 1 and vertex 4 is $P_{1,4}$.

The distance between vertex 1 and vertex 4 is:
\[
d(1,4)=3
\]
while distance between vertex 1 and vertex 5 is:
\[
d(1,5)=1
\]

\end{minipage}
}
\end{center}
\vspace*{1cm}
To describe the
importance of a node with respect to minimal paths in the graph, the
concept of betweenness helps. Betweenness (sometimes also referred as
{\em load}) for a given vertex is the number of shortest paths between any other nodes that traverse it. More formally, 
\begin{definition}[Betweenness]
 The {\em betweenness} $b(v)$ of vertex $v \in V$ is
\[
 b(v)=\sum_{v\neq s,t}\sigma_{st}(v)
\]
where $\sigma_{st}(v)$ is 1 if the shortest path between vertex s and vertex t goes through vertex v, 0 otherwise.
\end{definition}
\vspace*{1cm}
\begin{center}
\framebox[0.75\linewidth]
{
\begin{minipage}{0.70\linewidth}
\textbf{Example:}

Vertex 2 is involved in the following shortest paths:
\[
P_{1,3},P_{1,4},P_{3,1},P_{3,5},P_{4,1},P_{4,5},P_{5,3},P_{5,4}
\]
therefore betweenness of vertex 2 is:
\[
 b(2)=8
\]

\end{minipage}
}
\end{center}
\vspace*{1cm}
In the \PG domain however it is not always true that most central node are the most important nodes for the \G since they have to sustain the most number of shortest paths. It is not a priori true that the nodes that sustain the greatest electricity flow between two nodes in a real \PG are the one with highest betweenness since electricity flow could not follow the topological shortest path.
Although it is important to know what is the betweenness of the most important nodes, to shift from a local to a global graph measure it is also useful to have a high level picture of the state of betweenness for the whole graph. A statistic measure is then used:
\begin{definition}[Betweenness distribution]\label{def:betdist}
Consider the betweenness value $l$ of a node in a graph as a random variable, the
function 
\[
L_k=\{v\in G:\: b(v)=l\}
\]
is called {\em betweenness probability distribution} of the graph G.
\end{definition}
%
%
%
%
Another distinction that is available between graphs is their characterization as unweighted (as considered in Definition \ref{def:graph}) or weighted. The edges are not always the same, in fact, different importance can be associated to them considering their role in the network or physical properties connecting vertexes. From a formal point of view:
\begin{definition}[Weighted graph]
A {\em weighted graph} is a pair $G(V,E)$  where $V$ is the set of
vertexes and $E$ is the set of edges. An edge
$e_{i,j,w}=(v_i,v_j,w)$ is a triple where $v_i,v_j\in V$ and $w\in
\mathbb{R}$. $w$ is called {\em weight} of the edge.
\end{definition}
\vspace*{1cm}
\begin{center}
\framebox[0.75\linewidth]
{
\begin{minipage}{0.70\linewidth}
\textbf{Example:}

Figure~\ref{fig:weightedGraph} represents a weighted graph: each edge is characterized by a certain weight. The weight associated to the edge might be related to physical properties of the links (e.g., resistance of the cables in a Power Grid network) connecting the objects represented as vertexes in the graph.
\end{minipage}
}
\end{center}
\vspace*{1cm}
Specific properties  of a graph are best understood by resorting to the
matricial counterparts, in particular, to the Adjacency matrix and
Laplacian matrix graph representations.
\begin{definition}[Adjacency matrix]
The adjacency matrix $A=A(G)=(a_{i,j})$ of a graph $G$ of order $N$ is the $N \times N$ matrix given by
\[
 a_{ij} =
  \begin{cases}
   1      & \text{if } (v_i,v_j) \in E, \\
   0       & \text{otherwise.}
  \end{cases}
\]
\end{definition}
\begin{definition}[Laplacian matrix]
 Let $D=(D_{ij})$ be the $N \times N$ diagonal matrix with $D_{ii}=d(v_i)$ the degree of $v_i$ in $G$ and $A$ the adjacency matrix of $G$.\\
The matrix $L=D-A$ is the Laplacian matrix of graph $G$.
\end{definition}
\vspace*{1cm}
\begin{center}
\framebox[0.75\linewidth]
{
\begin{minipage}{0.70\linewidth}
\textbf{Example:}

The Adjacency matrix for graph $G$ is:
 \[
 A(G) =
 \begin{pmatrix}
  0 & 1 & 0 &0&1 \\
   1 & 0 & 1 &0&1  \\
   0 & 1 & 0 &1&0   \\
   0 & 0 & 1 &0&0\\
    1 & 1 & 0 &0&0  
 \end{pmatrix}
\]

The Laplacian matrix for graph $G$ is:
\[
 L(G) =
 \begin{pmatrix}
  2 & -1 & 0 &0&-1 \\
   -1 & 3 & -1 &0&-1  \\
   {}0 & -1 & {}2 &-1&{}0   \\
   {}0 & 0 & -1 &{}1&{}0\\
    -1 & -1 & {}0 &0&{}2  
 \end{pmatrix}
\]
\end{minipage}
}
\end{center}
\vspace*{1cm}


Another interesting property that is investigated for networks and
graphs is the so called \sw property. Although a complete coverage of
the \sw problem is beyond the scope of the present work, we recall the
basic definitions and refer to \cite{Watts98,Watts03}. We begin with the
clustering coefficient.
\begin{definition}[Clustering coefficient]
The {\em  clustering coefficient} of a node $v$ is
\[
\gamma_v=\frac{|E(\Gamma_v)|}{\binom{k_v}{2}} 
\]
where $|E(\Gamma_v)|$ is the number of edges in the neighborhood of
$v$ and $\binom{k_v}{2}$ is the total number of $possible$ edges in
$\Gamma_v$. The \emph{clustering coefficient} of graph $G$ is
$\gamma$, the average value of the clustering coefficient of all
nodes of the graph.
\end{definition}

\begin{definition}[Characteristic path length]
Let $v_i \in V$ be a vertex in graph $G$, the \emph{characteristic path length} for $G$, $L_{CP}$ is the median of ${d_{v_i}}$ where:
\[
 d_{v_i} = \frac{1}{(N-1)} \sum_{i\neq j}d(v_i,v_j)
\]
is the mean of the distances connecting $v_i$ to any other vertex $v_j$ in $G$ whose order is $N$.
\end{definition}
\vspace*{1cm}
\begin{center}
\framebox[0.75\linewidth]
{
\begin{minipage}{0.70\linewidth}
\textbf{Example:}

The clustering coefficient for vertex 1 is:
\[
 \gamma_1=1
\]
while for the entire graph it is the average of $ \{ \gamma_1=1, \gamma_2=\frac{1}{3}, \gamma_3=0, \gamma_4=0, \gamma_5=1 \} $ that is:
\[
 \gamma_G=0.467
\]

The characteristic path length is the median of
\[
\{ d_{v_1}=\frac{7}{4},d_{v_2}=\frac{5}{4},d_{v_3}=\frac{6}{4},d_{v_4}=\frac{9}{4},d_{v_5}=\frac{7}{4} \}
\]
that is:
\[
L_{CP}=\frac{7}{4}=1.75
\]

\end{minipage}
}
\end{center}
\vspace*{1cm}
\begin{definition}[Random Graph]
A graph $G(V,E)$ of a given size is a \emph{random graph} if it is the result of a
random process where the edges between any two nodes have been chosen independently with probability $p \in [0,1]$ among all possible $\binom{N}{2}$ edges.
\end{definition}
We remark that there exist several definitions of random graphs as
provided by Erd{\H{o}}s and R{\'{e}}nyi~\cite{Erdos1960},
Bollobas~\cite{bollobas01} or Durrett~\cite{Durrett06}. These mostly lead anyhow to the same asymptotic properties. This holds
also for the clustering coefficient e.g.~\cite{Newman2003} which
differs from the definition we proposed (in line with Watts and Strogatz in~\cite{Watts98} and Watts in~\cite{Watts03}). 

Small-world networks (SW), proposed by Watts and Strogatz in~\cite{Watts98},
own two important properties at the same time: the characteristic path length is
close in value to the one of a random graph (RG) ($CPL_{SW} \approx
CPL_{RG}$) and they have a much higher clustering coefficient ($CC_{SW} \gg
CC_{RG}$). Small-worlds are a better model than random graphs for social networks and
other phenomena~\cite{Milgram69,kochen89,Shi2008,Cancho2001} and thus
a model to keep in mind for the Power Grid, too.

Another investigation usually performed when analyzing \PG and that is almost always the motivation that drives \CNA studies related to electrical infrastructures is the investigation of reliability. Usually, the investigation involves evaluating the disruption behavior of the graph when its nodes or edges are removed. There are basically
two ways to perform this analysis: choosing the nodes to be removed randomly or
selecting the nodes following a certain property or
metric significant for the network. Commonly the metric used to remove nodes follows the  highest degree
or highest betweenness to simulate \emph{targeted attacks} that focus on specific nodes with certain properties or importance for the network.

\begin{figure}
\centering
\includegraphics[width=0.4\textwidth]{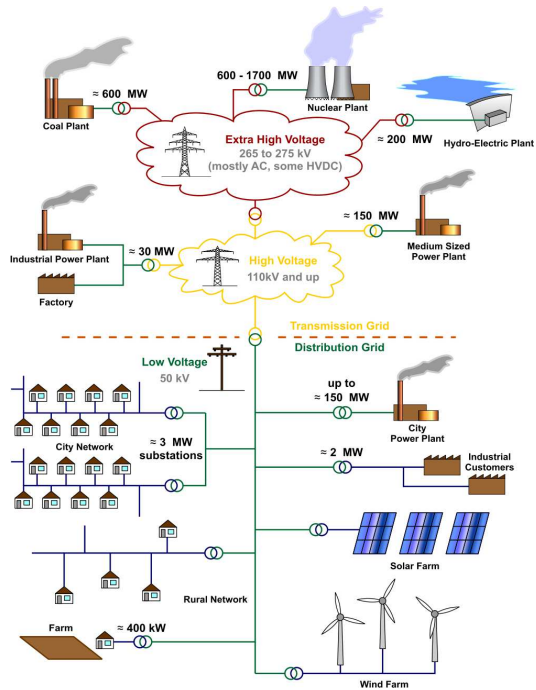}
 \caption{Organization of the \PG layers.}
 \label{fig:pglayers}
\end{figure}

Other terms to compare the various \PG studies involve more general
characteristics of the network under analysis. In particular, the
geographical location of the analyzed Grid is responsible for
topological properties due to the different morphological
characteristics of different countries. Another relevant aspect deals
with the layer of the \PG under investigation since differences can
emerge from a topological perspective investigating the different ends
in which the Grid is usually partitioned: High, Medium and Low
Voltage. An example on how the \PG is organized is shown in Figure~\ref{fig:pglayers}. It is also important to have information if the type of \PG graph under analysis comes from a real network infrastructure or it is a synthetic sample extracted from blueprint models for the \PG such as the Bus models of IEEE.

The motivations to include the works in this survey are based on the quality of the research performed, the rigor in the application of \CNA methodologies and the geography of the \PG analyzed in order to cover a broad spectrum of the infrastructure realized in the different countries and identify possible differences.

\section{The Power Grid as a Complex Network}\label{sec:relwks}

Complex network analysis studies are becoming more and more popular
given the amount of natural and human complex systems. The \PG is
clearly amenable to such studies and a number of these have been
performed on the \HV Grid. Here we describe the most important aspects
of each work under investigation. In particular, the works that are
considered in this review
are:~\cite{alb:str04,Crucitti04,Chassin05,holmgren06,Rosas-Casals2009,casals07,Corominas-murtra2008,Crucitti2005,Rosato2007,Watts03,PaganiAielloTSG2011,Mei11,casals10,bompard10,kim07,hines08,
wang10_2,nasiruz11,Han2011,wang10_c,Wang2009a,ding06,dwivendi10,dwivendi09,guohua08,sun05,pahwa10,kinney05,arianos09,
bompard09,hines10,pepyne07,Brummitt12}. These
have been chosen based on the following factors: they are specifically
about the Power Grid, they cover US, European, Chinese Grids or synthetic topologies of electrical engineering literature, they have
samples of different sizes and, most importantly, these are the
best-known and most representative works on the topic of CNA and Power Grid.


\subsection{Basic \PG characteristics}

The aspects considered in this first basic assessment of the studies take
into account general and non-technical aspects so to give a
global idea of the \G considered, see
Table~\ref{tab:compaBasicProp}. Several aspects of comparison are
considered: the number of nodes and lines composing the \G (second and
third column); the type of sample considered either a real  \G or
synthetic samples, for instance, coming from IEEE literature such as
IEEE Bus systems (fourth column); the type of \G analyzed (fifth
column) in belonging either to the transmission part (High Voltage) or
to the distribution part (Medium and Low Voltage); another essential
information deals with the geography of the \G (last column).

\begin{table}[htbp!]

\begin{center}
\begin{footnotesize}
    \begin{tabular}{| c| c |c| c |c | c |}
    \hline
    \textbf{Work} & \textbf{Number of} & \textbf{Number of}& \textbf{Sample }& \textbf{Network }& \textbf{Geography }\\ 
         &\textbf{Nodes }&\textbf{Lines}&\textbf{Type }& \textbf{Type}&\\\hline
    \cite{alb:str04} & $\sim$14000 & $\sim$19600 & Real & HV & North America\\ 
     \hline
    \cite{Crucitti04} & $\sim$300 & $\sim$500 & Real& HV & Italy \\ 
    \hline
    \cite{Chassin05}  & $\sim$314000&N.A. & Real& HV&  North America  \\ 
    \hline
    \cite{holmgren06}& $\sim$4800& $\sim$5500& Real& HV & Scandinavia \\ 
      \hline
    \cite{Rosas-Casals2009}  & $\sim$2700&$\sim$3300 & Real& HV &Europe \\ 
     \hline
    \cite{casals07} & $\sim$3000&$\sim$3800 & Real & HV&Europe \\ 
   \hline
    \cite{Corominas-murtra2008}  & $\sim$3000&$\sim$3800 & Real& HV&Europe \\ 
     \hline
    \cite{Crucitti2005}  & $\sim$370 & $\sim$570& Real& HV&Italy, France     \\ 
    &&&&&and Spain \\ \hline
    \cite{Rosato2007}  & $\sim$370& $\sim$570& Real & HV &Italy, France  \\ 
     &&&&&and Spain\\ \hline
   \cite{Watts03} & $\sim$4900& $\sim$6600& Real & HV &Western US\\ 
     \hline
    \cite{Wang2010} & $\sim$8500& $\sim$13900& Synthetic & HV&Western US and \\ 
    & &&and real& & New York State Area\\ \hline
    \cite{PaganiAielloTSG2011} & $\sim$4850& $\sim$5300& Real&MV/LV & Netherlands \\
     \hline
    \cite{Mei11}\footnotemark[1]
& $\sim$210& $\sim$320& Synthetic& HV & China \\ 
    & &&and real& & \\ \hline
   \cite{casals10} & N.A.&N.A. & Real & HV&Europe \\ \hline
   \cite{bompard10} & 300& 411 & Synthetic & HV&  \\ \hline
   \cite{kim07} & $\sim$6400& $\sim$8700 & Synthetic & HV& North America,  \\ 
    & &&and real&& Scandinavia and Korea\\ \hline
    \cite{hines08} & 300& 411 & Synthetic & HV& \\ \hline
    \cite{wang10_2} & $\sim$8500& $\sim$13900& Synthetic & HV&Western US and \\ 
    & &&and real& & New York State Area\\ \hline
    \cite{nasiruz11} & $\sim$30& $\sim$13900& Synthetic & HV&Western US and \\ 
    & &&and real& & New York State Area\\ \hline
    \cite{Han2011}& $\sim$900&$\sim$1150&Real&HV&China\\ \hline
    \cite{wang10_c} & $\sim$3200& $\sim$7000& Synthetic & HV& New York State Area\\ 
    & &&and real& & \\ \hline
        \cite{Wang2009a}& $\sim$4900&$\sim$6600&Real&HV&Western US\\ \hline
        \cite{ding06}& $\sim$1700&$\sim$1800&Real&HV&China\\ \hline
    \cite{dwivendi10}& $\sim$39&$\sim$46&Syntethic&HV&\\ \hline
    \cite{dwivendi09}& $\sim$39&$\sim$46&Syntethic&HV&\\ \hline
    \cite{guohua08}& $\sim$2500&$\sim$2900&Real&HV&China\\ \hline
    \cite{sun05}& $\sim$15400&$\sim$18400&Real&HV&North America and China\\ \hline
     \cite{pahwa10}& $\sim$550&$\sim$800&Synthetic&HV&\\ \hline
    \cite{kinney05} & $\sim$14000 & $\sim$19600 & Real & HV & North America\\ \hline
    \cite{arianos09} & $\sim$90 & $\sim$120 & Synthetic & HV & \\ \hline
    \cite{bompard09} & $\sim$550 & $\sim$700 & Synthetic & HV & Italy\\ 
    & &&and real& & \\ \hline
    \cite{hines10} & $\sim$29500 & $\sim$50000 & Synthetic & HV & North America\\ 
    & &&and real& & \\ \hline
    \cite{pepyne07} &  $\sim$400 & $\sim$700 & Synthetic & &\\ \hline
    \cite{Brummitt12} & $\sim$900 & $\sim$1300 & Synthetic & HV & South-East US\\ 
     & &&and real& & \\ \hline
    \cite{chaitanya11} & $\sim$60 & $\sim$110 & Real& HV & India\\ \hline

    \end{tabular}
    \end{footnotesize}
\end{center}

\caption{Comparison between studies using CNA for the Power Grid.\label{tab:compaBasicProp}}
\end{table}

\footnotetext[1]{The values for nodes and lines in this table refer only to a snapshot of Shanghai Power Grid}

Albert \etal~\cite{alb:str04} study the reliability aspects of the United States Power Grid. They build a graph based on the information of the POWERmap system (developed by Platts company) consisting of 14099 nodes representing power plants and substations and 19657 edges representing \HV lines (115-765 kV). This same \PG sample is used also in the study by Kinney \etal~\cite{kinney05} where the reliability of the \G is analyzed with a different set of metric compared to those used by Albert \etal.

Crucitti \etal~\cite{Crucitti04} analyze the Italian \HV \PG from a topological perspective. They build a model based on data from GRTN (the Italian Grid manager at that time) consisting of 341 substations (nodes) and 517 transmission lines (edges) belonging to the \HV segment (220-380 kV). 

Chassin \etal~\cite{Chassin05} analyze the North American Power Grid. They treat the North American Grid as composed by two different networks, i.e., the Eastern Grid and the Western Grid, that are analyzed separately. This assumption is justified by the authors by the presence of a small linking between the two Grids (this is realized on purpose to avoid the spreading of blackouts across the entire country) realized in direct current technology. In addition, the data they use contain much more nodes and links (almost 236000 nodes for the Eastern and more than 78000 for the Western Grid, and it is based on the estimations of typical load and maximal capacity of distribution electrical feeders); data come form Western Electricity Coordinating Council (WECC) and North American Electric Reliability Council (NERC), respectively. 

Holmgren~\cite{holmgren06} analyzes the Nordic \PG involving the \HV Grids of Sweden, Finland, Norway, and the main part of Denmark which give an overall graph composed by almost 4800 nodes and more than 5500 edges. The author compares the Nordic network with the Western U.S. Power Grid. 

Casals \etal~\cite{Rosas-Casals2009} analyze the whole European \PG and try to extract non-topological reliability measures investigating the topological properties of the network. The \PG analyzed is the \HV end composed of almost 2800 nodes that span across all European continent. Casals \etal~\cite{casals07} consider the \HV Grids of many European countries analyzing them together and as separate entities having an overall sample of more than 3000 nodes and around 4300 edges. Sol\'{e} \etal~\cite{Corominas-murtra2008} go further in exploring the same \PG data analyzed in~\cite{casals07}, in particular, they focus on analyzing the targeted attacks to European Power Grids. Rosas-Casals~\cite{casals10} once again considers the European \HV \PG to understand the distribution of failures that have characterized it from 2002 to 2008.

Crucitti \etal~\cite{Crucitti2005} analyze the \HV \PG of Italy (127 substations and 171 lines belonging to 380kV network), France (146 substations and 223 lines belonging to 400kV network) and Spain (98 substations and 175 lines belonging to 400kV network).   This same sample is analyzed by Rosato \etal~\cite{Rosato2007} to investigate the main topological properties of these Grids. 

Watts~\cite{Watts03} dedicates a subsection to explore the properties of the Western States Power Grid of the U.S. This \G is treated as an undirected unweighted graph in which all the nodes of the network no matter their task (e.g., generators, transformers, substations) are equally considered as nodes. The same assumption is done for the edges: the transmission lines are considered equal even if the voltages they involve can be extremely different (the Grid considered belongs only to \HV segment with lines varying from 345 to 1500 kV). The overall graph is quite large (4941 nodes). This same network is the base for the study by Wang~\etal~\cite{Wang2009a} which focuses on the vulnerability of the Grid.

Wang \etal~\cite{Wang2010,wang10_2} investigate both on real \PG samples for about 8000 nodes (the networks analyzed are the American NYISO and WSCC) and 4 synthetic reference models belonging to the IEEE literature that account for about 500 nodes. In their other work,~\cite{wang10_c}, Wang \etal~consider once again the NYSO-2935 Bus and the synthetic IEEE model: IEEE 300-Bus system.

The work of Bompard \etal~\cite{bompard10} refers to the IEEE Bus blueprint infrastructure which applies an enhanced betweenness analysis to the IEEE 300-Bus. Similarly the work of Dwivedi \etal~\cite{dwivendi10,dwivendi09} which study respectively the IEEE 39-Bus and the IEEE 118-Bus. Also the study of  Pahwa \etal~\cite{pahwa10} refers to the synthetic Buses from IEEE (IEEE 300-Bus and IEEE 118-Bus) and the WSCC 179-Bus equivalent. Arianos \etal~\cite{arianos09} study smaller IEEE networks: the IEEE 30-Bus and the IEEE 57-Bus.

In our work~\cite{PaganiAielloTSG2011}, we study the \MV and \LV end of the \PG with special focus on the northern Netherlands situation. The overall sample is almost 700 nodes for the \LV part and 4200 for the \MV one.

Mei \etal~in their book~\cite{Mei11} perform various simulation related to \PG vulnerability and blackout conditions considering the synthetic models of IEEE literature (e.g., IEEE 14-Bus, IEEE 30-Bus, IEEE 39-Bus and IEEE 118-Bus) and also real samples of Chinese Power Grid. In particular they analyze the 500kV Northeast \PG of China and a snapshot of Shanghai \HV \PG which consist of more than 200 nodes and 300 lines.

Kim \etal~\cite{kim07} analyze basic metrics of several Grids both the synthetic ones belonging to IEEE literature such as 14-Bus and 118-Bus (respectively 14 nodes and 20 edges and 118 nodes and 179 edges) and real \HV networks. The real samples are the American Mid-Continental Area Power Pool (MAPP) of 230kV and higher composed of 575 nodes and 754 edges, the Electric Reliability Council of Texas (ERCOT) of 345kV composed of 148 nodes and 209 edges and the Western Systems Coordinating Council (WSCC) from 100kV and above which has about 4600 nodes and more than 6200 edges. In addition, they analyze the Nordel network in Scandinavia (100kV and above) characterized by 410 nodes and 564 edges and the Korea Electric Power Corporation (KEPCO) (66kV and above) made of 553 nodes and 783 edges.

Hines \etal~\cite{hines08} uses the \HV synthetic model IEEE 300-Bus system as the reference network on which compute centrality metrics exploiting the impedence representation of lines connecting Buses (nodes).

Han \etal~\cite{Han2011} investigate the topological characteristics of two Power Grids sections belonging to the Chinese \HV network: the Anhui \PG (84 nodes and 112 edges) and the bigger East China \PG (769 nodes and 1029 edges). Also focused on the Chinese \HV Grid is the work of Ding \etal~\cite{ding06} analyze the topological properties of two samples of the Chinese \HV Power Grid. In particular, the Sichuan-Chongqing \G (724 nodes and 771 edges) and the Guangdong \G (956 nodes and 1049 edges).
Guohua \etal~\cite{guohua08} analyze another sample of the Chinese Power Grid, in particular the North China \PG which is composed by 2556 nodes and 2892 edges.

In one of the first comparison of power Grids~\cite{sun05}, Sun compares the Western American Grid ($\approx$5000 nodes and $\approx$6600 edges), the North China \PG ($\approx$8000 nodes and $\approx$9000 edges) and the Center China \PG ($\approx$2400 nodes and $\approx$2800 edges). The model used considers the transmission lines (above 110kV for the Chinese samples and above 115kV for the American one) as unweighted and undirected edges, while generators, substations and transformers are considered all equal nodes of the graph.

The Italian transmission is once again the example of network used to test \PG vulnerabilities~\cite{bompard09}, comprising 521 nodes and 679 transmission lines. In addition in the same work a synthetic Bus with 34 nodes is tested as well.

The work of Hines \etal~\cite{hines10} investigates the vulnerability aspects both of synthetic samples (IEEE 300-Bus) and 40 control areas (out of the 136 areas) of the North American Eastern Interconnect \G which represent about 29000 nodes and 50000 transmission lines of the Grid.

The work of Pepyne~\cite{pepyne07} focuses on evaluating the cascading effects of synthetic models  coming from the IEEE literature (e.g., IEEE 57-Bus and IEEE 118-Bus) and on a sample satisfying the \sw model~\cite{Watts98,Watts03} of 200 nodes and 400 edges.

Brummitt \etal~\cite{Brummitt12} investigate the cascading effects in interdependent networks (i.e., networks which are loosely coupled) and the effects of adding connectivity (e.g., power lines) to mitigate or aggravate cascade effects. For this purpose two loosely connected samples of the South-East region of the United States are used as samples, for a total of about 900 nodes and 1300 edges.

A new geography is explored by Chaitanya \etal~\cite{chaitanya11} who investigate the properties of an \HV \G of the eastern region of India. The network is built on 63 nodes and 113 transmission lines.

Notice that the numbers in the second and third column are not the exact numbers, but they are an approximation to give the idea of the importance of the sample.

\begin{figure}[h!]
 \centering
\includegraphics[width=0.5\textwidth]{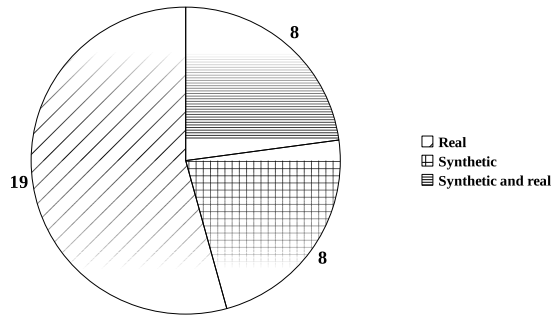}
 \caption{Number of studies that consider real \PG samples or synthetic models.}
 \label{fig:realSynth}
\end{figure}

\begin{figure}[h!]
 \centering
\includegraphics[width=0.5\textwidth]{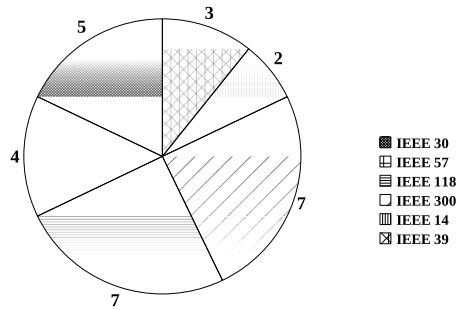}
 \caption{IEEE literature bus model analyzed.}
 \label{fig:ieeebus}
\end{figure}

\begin{figure}[h!]
 \centering
\includegraphics[width=0.5\textwidth]{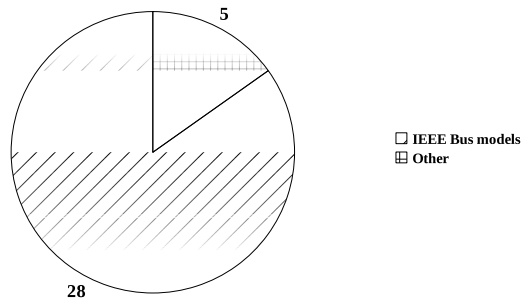}
 \caption{Number of models used coming from the IEEE literature compared to other models.}
 \label{fig:busOther}
\end{figure}

\begin{figure}[h!]
 \centering
\includegraphics[width=0.7\textwidth]{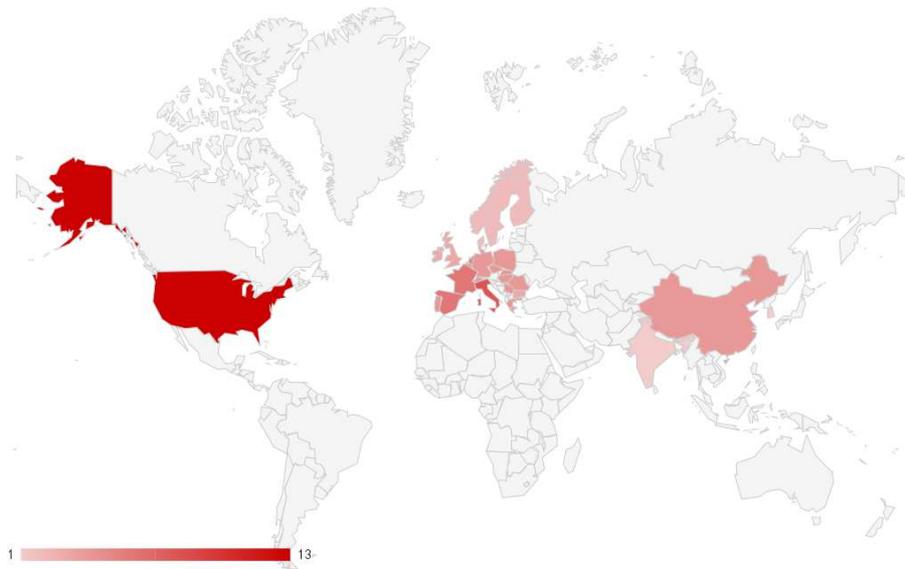}
 \caption{Map of the \PG infrastructure studied using CNA approach.}
 \label{fig:world}
\end{figure}

\begin{figure}[h!]
 \centering
\includegraphics[width=0.5\textwidth]{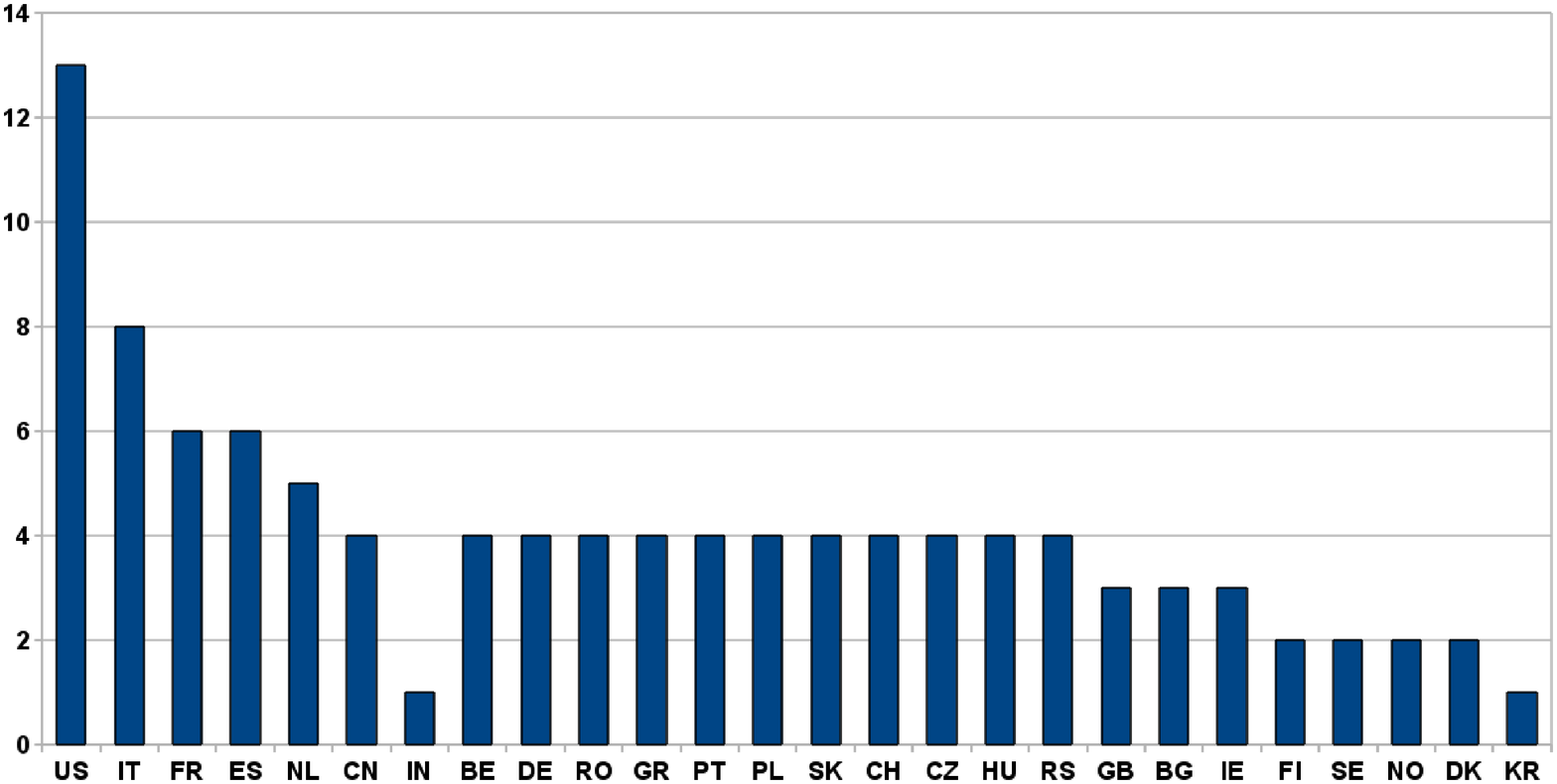}
 \caption{Number of times \PG of a given country are investigated.}
 \label{fig:country}
\end{figure}

\vspace*{1cm}
\begin{center}
\framebox[0.75\linewidth]
{
\begin{minipage}{0.70\linewidth}
\textbf{Summary:}

\noindent
In summary, the data are in the most cases extracted from real samples, that is, they represent real electric infrastructures deployed other works in addition to real Power Grids while fewer studies consider synthetic models as shown in Figure~\ref{fig:realSynth}. Most of these synthetic approaches consider samples coming from IEEE blueprints such as IEEE Bus systems (a representation of the various IEEE Bus models used in the surveyed articles is shown in Figure~\ref{fig:ieeebus}), while very few concentrate only on other synthetic samples (e.g., non-IEEE models, \sw models, random graphs); the number of synthetic model used is shown in Figure~\ref{fig:busOther}. Almost all samples belong to the \HV end of the \PG that are the lines used for long range transmission to which big power plants are attached too; the only exception is our study~\cite{PaganiAielloTSG2011} that is focused on the distribution part of the Grid (i.e., \MLV network). From a geographical perspective the samples are mainly localized in the United States or in Europe with some studies that consider Chinese \HV samples; a map of the countries whose Grids are analyzed is represented in Figure~\ref{fig:world} and the number \G analyzed for a given country is shown in Figure~\ref{fig:country}. Another main commonality is to treat the \G as an undirected graph where each substation or transformer represents a node and each line transporting electricity is an edge.

\end{minipage}
}
\end{center}

\subsection{Statistical global graph properties} \label{sec:statProp}

The main characteristics from a graph and \CNA perspective of the \Gs under analysis are summarized in Table \ref{tab:compaGridGraphs}. Several aspects of comparison are considered: the \textit{order} ($N$) and \textit{size} ($M$) of the graph (second and third column) corresponds to the number of nodes (\textit{order}) and number of lines (\textit{size}) actually in the Power Grids. The average degree, computed as $<\!k\!>=\frac{2M}{N}$, gives a general idea of how many vertexes is an average vertex connected to (fourth column). Fifth, sixth and seventh column give information about the type of statistical analysis performed on the graph, in particular, the assessment of node degree distribution and betweenness distribution together with an evaluation of the path length are considered. Another term of comparison deals with the type of graph analyzed taking into account weights or simply use the unweighted definition of graph. Last two columns of the table consider the type of aim of the graph analysis either an investigation of the disruption behavior of the graph or the evaluation of the \sw properties.

\begin{sidewaystable}[htpb!]

\begin{center}
\begin{footnotesize}
    \begin{tabular}{| c| c |c| c |c | c |c |c | c | c |}
    \hline
    \textbf{Work }& \textbf{Sample }& \textbf{Sample}&\textbf{Average }& \textbf{Node Degree  }& \textbf{Betweenness }&\textbf{Path Length}& \textbf{Weighted/Unweighted }&\textbf{ Resilience  }&\textbf{ Small-world }\\ 
         &\textbf{Order} &\textbf{Size }&\textbf{degree}& \textbf{Distribution }& \textbf{Distribution }& \textbf{Analysis}&\textbf{Analysis }& \textbf{Analysis }&  \textbf{Investigation}\\
     & && & \textbf{Statistics} &\textbf{Statistics}& &  &&\\ \hline
   \cite{alb:str04}& $\sim$14000 & $\sim$19600& $\sim$2.80 & \checkmark &\checkmark & & Unweighted & \checkmark & \\ \hline
    \cite{Crucitti04} & $\sim$300 & $\sim$500& $\sim$3.33 & \checkmark &\checkmark &Indirectly through &Weighted not based  & \checkmark & \\ 
     & & & & &&efficiency metric&on physical properties & & \\\hline
    \cite{Chassin05}  & $\sim$314000&N.A. &N.A.& \checkmark & &  &Unweighted & \checkmark &  \\  \hline
   \cite{holmgren06}& $\sim$4800& $\sim$5500 &  $\sim$2.29& \checkmark & &\checkmark &Unweighted & \checkmark & \checkmark \\ 
      \hline
    \cite{Rosas-Casals2009} & $\sim$2700&$\sim$3300& $\sim$2.44 & \checkmark & & & Unweighted  & \checkmark & \\ 
    \hline
   \cite{casals07} & $\sim$3000&$\sim$3800 & $\sim$2.53&  \checkmark &&  \checkmark &Unweighted & \checkmark & \checkmark \\ 
    \hline
    \cite{Corominas-murtra2008}  & $\sim$3000&$\sim$3800& $\sim$2.53 & \checkmark & & & Unweighted & \checkmark & \\ 
     \hline
   \cite{Crucitti2005} & $\sim$370 & $\sim$570& $\sim$3.08 &  & &Indirectly through &  Unweighted & \checkmark & \\ 
   &&&&&& efficiency metric&&&\\ \hline
    \cite{Rosato2007}  & $\sim$370& $\sim$570& $\sim$3.08&  \checkmark &&\checkmark &Unweighted  & \checkmark & \\ 
     \hline
    \cite{Watts03}  & $\sim$4900& $\sim$6600& $\sim$2.69&  & &\checkmark &Unweighted   & & \checkmark\\ 
     \hline
    \cite{Wang2010} & $\sim$8500& $\sim$13900& $\sim$3.27& \checkmark &  &\checkmark &Unweighted   &  & \checkmark\\ 
    & &&&&&&and impedance analysis& &\\ \hline
    \cite{PaganiAielloTSG2011} & $\sim$4850& $\sim$5300& $\sim$2.18 & \checkmark & \checkmark & \checkmark & Both & \checkmark & \checkmark \\
   \hline
   \cite{Mei11}\footnotemark[2]
   & $\sim$210& $\sim$320& $\sim$3.05 &  &  & \checkmark & Both & \checkmark & \checkmark \\
   \hline
    \cite{casals10} & N.A.& N.A.& &  &  & &  & \checkmark &  \\
   \hline
   \cite{bompard10} & 300& 411&2.74 &  &  & & Both & \checkmark &  \\
   \hline
   \cite{kim07} & $\sim$6400& $\sim$8700 &2.72 &  &  &\checkmark & Unweighted & \checkmark & \checkmark \\
   \hline
   \cite{hines08} & 300& 411 &2.74 & \checkmark (chart only) & \checkmark & & Both & \checkmark &  \\
   \hline
   \cite{wang10_2} & $\sim$8500& $\sim$13900& $\sim$3.27& \checkmark &  &&Unweighted   & \checkmark & \\ \hline
   \cite{Han2011} & $\sim$900&$\sim$1150& $\sim$2.55& \checkmark & \checkmark &\checkmark &Weighted  & \checkmark & \checkmark \\ \hline
   \cite{wang10_c} & $\sim$3200&$\sim$7000& $\sim$4.375& \checkmark (chart only)& \checkmark & &Weighted  & \checkmark &  \\ \hline
   \cite{Wang2009a}  & $\sim$4900& $\sim$6600& $\sim$2.69&  & & &Unweighted   & \checkmark & \\ 
     \hline
      \cite{ding06} & $\sim$1700&$\sim$1800& $\sim$2.12&  &  &\checkmark &Both  &  & \checkmark \\ \hline
   \cite{dwivendi10} & $\sim$39&$\sim$46& $\sim$2.36&  &  & &Weighted  & \checkmark &  \\ \hline
    \cite{dwivendi09} & $\sim$150&$\sim$46& $\sim$2.36&  &  & &Weighted  & \checkmark &  \\ \hline
	\cite{guohua08}& $\sim$2556&$\sim$2892&$\sim$2.26&  &  & &Weighted  & \checkmark &  \\ \hline
	\cite{sun05}& $\sim$15400&$\sim$18368&$\sim$2.39 &\checkmark (for one sample only) &  &\checkmark &Unweighted  & \checkmark & \checkmark \\ \hline
	\cite{pahwa10}& $\sim$550&$\sim$800&$\sim$2.91 &&  &\checkmark &Unweighted  & \checkmark & \\ \hline
	\cite{kinney05}& $\sim$14000 & $\sim$19600& $\sim$2.80 &  & & & Weighted not based on & \checkmark & \\ 
	 & & & & &&&on physical properties & & \\\hline
	 \cite{arianos09}& $\sim$90&$\sim$120&$\sim$2.67 &&  & &Weighted  & \checkmark & \\ \hline
	\cite{bompard09}& $\sim$550&$\sim$700&$\sim$2.55 &&  & &Weighted  & \checkmark & \\ \hline
	\cite{hines10}& $\sim$29500&$\sim$50000&$\sim$3.39 &&  & \checkmark &Weighted  & \checkmark & \\ \hline
	\cite{pepyne07} &  $\sim$400 & $\sim$700& $\sim$3.5 &&  &\checkmark  &Weighted  & \checkmark & \checkmark \\ \hline
	\cite{Brummitt12} & $\sim$900 & $\sim$1300 &$\sim$2.89 &&  & &Unweighted  & \checkmark & \\ \hline
	\cite{chaitanya11}& $\sim$60&$\sim$110&$\sim$3.67 &&  & &Unweighted  & & \\ \hline
    \end{tabular}
    \end{footnotesize}
\end{center}

\caption{Comparison of the main characteristics of the graphs related to Power Grids.\label{tab:compaGridGraphs}}
\end{sidewaystable}

\footnotetext[2]{The values for nodes and lines in this table refer only to a snapshot of Shanghai Power Grid.} 

Many studies \cite{PaganiAielloTSG2011,Wang2010,Watts03,Corominas-murtra2008,casals10} remark the limited value of the average node degree, generally between 2 and 3, for the \PG especially if compared to other types of Complex Networks (e.g., the Web, social networks). This is due to the physical, geographical and economical constraints that are associated to the substations and power cables. 

Half of the studies focuses on the investigation of the node degree distribution statistics since this information is a key to find what kind of theoretical probability model is beyond the sample, allowing to establish which kind of network the sample can be associated with. The other studies that do not take into account node degree distribution statistics focus on other very specific aspects of the Complex Network Analysis. Betweenness distribution statistics are less common, in fact, only few studies dig into this property, although it provides essential information related to the load sustained by the nodes of the network. 

Almost half the studies take into account the path length to investigate the effort that it takes to move from one node to any other one. The study of the path length is usually not performed per se, but it is essential to then proceed in the investigation of the \sw property of the network. Almost all the studies that investigate path properties then go further and evaluate the \sw characteristics as well.  

Half of the studies performs analysis considering the unweighted definition of graph, while the other half takes into account weights for investigating the graph. Crucitti \etal~\cite{Crucitti04} use a weighted graph that anyway has no relationship with the physical properties of the considered Power Grid. The weight used is related to the betweenness managed by a node and it is partitioned between the edges it is connected to. Wang \etal~\cite{Wang2010} consider an impedance analysis therefore dealing with the physical properties of the lines, but the \PG graph are then not considered with the weighted definition, but only a probability distribution of the impedance is computed. Again Wang \etal~\cite{wang10_c} use a weighted representation of the \G by using an admittance matrix which is based on the Laplacian matrix representation of the \PG enriched with the information about the admittance of the lines connecting the nodes. In~\cite{PaganiAielloTSG2011}, we perform a weighted analysis considering the resistance of cables as weights for the edges in the graph and we compute the same set of statistics for the unweighted and weighted definition of the graph. One reason for the lack of weighted \CNA analysis on real samples of the \PG is probably due to the difficulty, first of having Grid data, and second of having the detailed information of the cables involved. The work of Mei \etal~\cite{Mei11} is particularly rich and complete in the type and characteristics of the graph that are considered. Several models are considered based on the specific aspect of the \PG that are analyzed in the various sections of the book. For instance the evaluation of the \sw properties of the \PG and the power flow characterizing the \G the graph is built considering the admittance values that characterize the lines. In another chapter which analyzes the vulnerability when considering the static behavior of the \PG the authors build a a directed graph with weights on the edges that represent the length of the lines. Again in another chapter describing the growth and evolution models of the \PG the graphs generated are considered undirected and unweighted. The weights considered in~\cite{bompard10} represent the power flows that move through the lines. Weights are also considered in~\cite{hines08} where an impedance matrix representing the electrical properties of the physical network is used to assign weights to edges in the IEEE Bus structure used. In~\cite{Han2011} weights are represented by the reactance of the transmission lines which are assigned as the values of the adjacency matrix representing the transmission lines. Another study that uses weights assigned to the edges representing the impedence of the transmission line is~\cite{ding06} where two samples of the \HV Chinese \G are analyzed. Other examples that take into account weights are~\cite{dwivendi10,dwivendi09} where the the graph representing the network is considered directed, since the \PG is assumed in a steady state, and the weights are represented by the admittance matrix characterizing the lines.
Guohua \etal~\cite{guohua08} analyze the North China \PG and the graph model that is used considers a weighted and directed graph, associating weights both on nodes (amount of input power in the node) and edges (power transmitted on the transmission line). 
The work of Sun~\cite{sun05} is mainly a comparison of the first results obtained considering \CNA techniques applied to the \PG field for the American \PG and two samples of the Chinese \G (North and Center China \PG). 
The analysis performed by Kinney~\etal~consider once again the U.S. Power Grid used in~\cite{alb:str04}, but this time the graph is considered weighted both for edges and for nodes. In particular a value of efficiency in the range of [0,1] is assigned to the edges representing their ability to transmit power; for nodes a value of capacity is provided which is proportional to the betweenness of the same node.
Arianos \etal~\cite{arianos09} evaluate the vulnerability of two samples coming from the IEEE Bus literature (IEEE 30-Bus and IEEE 57-Bus). In their analysis the authors use impedences values and power distribution factor associated to the transmission lines to have a model closer in his behavior  to the real electric flow. The model of the electrical flows considered is a direct current model which avoids much of the complexities behind the real energy flow exchanges, but that in a \CNA study is sufficient to have a good idea and approximation of the power flows in the network. This same approach of characterizing transmission lines with their physical parameters representing impedences and power distribution factor is used by Bompard \etal~\cite{bompard09} where the vulnerability results obtained with just topological measures are compared with a model that takes into account enhanced physical aspects as well is applied both to a synthetic network (a Bus with 34 nodes) and to the Italian \HV Grid.
Hines \etal~\cite{hines10} consider for their reliability evaluation the IEEE 300-Bus and a considerable sample of the North American Eastern Interconnect  (almost 30000 nodes and 50000 edges). Their aim is to evaluate the adequacy of pure topological measures compared to measures that  take into account also the electrical and physical properties of the Grid. Therefore, the authors consider a simplification of the Grid that exploits the direct current model for calculating the power flow in the network. To build this model of power flow physical/electrical properties are added to the pure topological description of the Grid. In particular, reactance values are assigned to the links and at each node a voltage phase angle is attributed so that the power flow flowing through a transmission line can be computed knowing the amount of power flowing through the neighboring nodes.
The study of Pepyne~\cite{pepyne07} focuses on the investigation of cascading effects in IEEE model buses (i.e., IEEE 57-Bus and IEEE 118-Bus) and on a \sw based network (200 nodes and 400 edges) using a DC power flow model to represent power flows to assess the lines affected in a cascading situation.
Brummitt~\etal~\cite{Brummitt12} consider a small section of the South-East US \G in order to evaluate the effects of enhanced connectivity of networks that are loosely connected. The two samples are about 400 and 500 nodes and they are connected with each other by just eight edges.
The Indian \PG is explored in~\cite{chaitanya11} where the very basic properties (e.g., average node degree, clustering coefficient) of the directed graph used to represent the eastern region Indian \HV \PG are investigated.

\vspace*{1cm}
\begin{center}
\framebox[0.75\linewidth]
{
\begin{minipage}{0.70\linewidth}
\textbf{Summary:}

\noindent
From Table~\ref{tab:compaGridGraphs} a difference appears: the studies closer to a topological characterization uses unweighted representation of the edges of the \G and consider always the node degree distribution in the analysis, since it is an important element to define the type of network under study (e.g., scale-free network). On the other hand, the studies that apply a weighted representation of the graph do not consider the node degree distribution statistics neither considering the unweighted definition of node degree nor using a definition that takes into account weights as proposed in~\cite{Newman04}. This last aspect might worth to be considered since the degree distribution properties of the network might change the picture of the node degree distribution in comparison with unweighted studies.

Centrality measures are not often used, exceptions are~\cite{Mei11,wang10_c,bompard10,dwivendi10} and the few other that compute between distribution statistics, to identify the statistical distribution of critical nodes. More attention to centrality measures especially using weighted representation of the \PG graphs or models that provide the capacity or energy flows through the \G might be beneficial in understanding the most critical nodes or lines in the power system.
Another recurring theme in the \CNA involving the \PG is the reliability analysis, and actually it is the main motivation that drives these kind of studies. In fact many works were performed after major blackout occurred, such as the  North
  American black-out of
  2003\footnotemark[2]
  or the Italian one of
  2003\footnotemark[3] (e.g.,~\cite{Albert04,Crucitti2004a,Crucitti04,Chassin05}) or anyway mention blackouts as the main motivation for the work. 
The fragility and resilience properties of the \PG has been the major reason of concern that has
determined the focus of such \CNA studies on the
\HV network. In fact almost all studies consider the behavior of the Grid to various attacks to its nodes or edges.

\end{minipage}
}
\end{center}
\footnotetext[3]{\url{http://news.bbc.co.uk/2/hi/americas/3152451.stm}}
\footnotetext[4]{\url{http://news.bbc.co.uk/2/hi/3146136.stm}}

\subsection{ The \sw property}\label{sec:sw}

The \sw property in network has received lots of attention starting with sociological studies~\cite{milgram67,Milgram69}, but more recently with application of this concept and model to many more classes of networks~\cite{Watts98,Watts03,Strogatz2001,Kleinberg00}.
Among the studies analyzed \sw property investigation is performed by ten out of the thirty two. The various studies look for the satisfaction of the \sw property described by Watts~\cite{Watts03} then together with Strogatz~\cite{Watts98}. 

Holmgren~\cite{holmgren06} performs a comparison of the Nordic network with a random graph with the same number of nodes and edges. The results show an average path length for the sample double compared to the random graph, but a clustering coefficient almost one order of magnitude bigger than the random graph one, so the author concludes the Nordic Grid belongs to the small-world class network.

Casals \etal~\cite{casals07} perform a comparison between each sample of the European \G and random graphs. The results show that the majority of samples satisfy the small-world conditions.

Watts~\cite{Watts03} dedicates his book to illustrating the \sw phenomenon and a section is dedicated to the analysis of the Western States Power Grid of the American network. He notes that sparseness of the graph corresponding to the \G violates one of the assumption usually necessary to have a small-world. Compared to other types of networks (e.g., social network of actors' interactions in movies), the clustering coefficient is quite small and the characteristic path length quite big, however, despite these characteristics, he states that the \sw property holds for the Western United States Power Grid. 
A statistic that emerges is the high fraction of edges that are also \emph{shortcuts} (i.e., if the edge is removed the shortest path between the same edge's ends is increased more than 2) for the graph which is around 80\%. It might be due to the sparseness of the graph which is implied by physical and economical limitations of adding lines to substations. Watts also notices that also the way the \G has developed supports the \sw concept: many independent and disconnected Grids have been connected together with the aim of sharing and exchanging power excess between remote locations, enhancing reliability and efficiency. Watts also notices that the model underlying the Western States Power Grid is closer and better explained quantitatively from a \emph{relational model}
than a \emph{dimensional model}. 
The former model consider the creation of edges as a function of the preexisting edges in the graph, as if the previous relationships between nodes were to a certain extent kept. The latter, on the other hand, considers the creation of edges as a function of the particular spatial location of the vertex and the physical distance to another vertex.
The better explication by a relational model is quite surprising and counter-intuitive, but can be justified by the inability of the dimensional model to admit occasional global edges spanning across nodes otherwise very far apart. This last property is the key to keep the small-world characterization.

Wang \etal~\cite{Wang2010} also investigate the \sw properties for their samples and they state that the model proposed by Watts and Strogatz \cite{Watts98}, is only able to capture some features of the Power Grid, since the \PG is sparsely connected compared to small-world networks. They also notice that the basic condition required by Watts and Strogatz's model is not satisfied by the \PGs under test.

In our study~\cite{PaganiAielloTSG2011}, we perform an investigation about the \sw properties of the \MLV network comparing the sample topologies with random graphs with same order and size. The results show that this end of the network seems even less close to \sw properties than the \HV due to a general very small clustering coefficient.

In~\cite{Mei11} the \sw concept is an essential aspect across the whole book. Actually the authors take as an assumption the fact that the \HV \PG is a \sw network. This is justified by the authors considering the type of evolution process that the \G is subject to in its development. In fact, it consists in a higher connectivity in a ``local-world'' where nodes (substations and power plants) are physically closer, therefore forming tighter connections locally, and only few long distance links that help in reducing the average path length of the network. In the \sw model also load and physical parameters are applied to make the model closer to the real \PG system. This model is analyzed to investigate the vulnerability of the network and its disruption behavior.

Kim \etal~\cite{kim07} investigate the \sw property of several \HV samples in particular comparing the metrics of characteristic path length and clustering coefficient to the theoretical ones for random graphs. Many of the samples (MAPP, Nordel, KEPCO and WSCC networks) satisfy the \sw conditions. The authors also define a so called ``small world-ness index'' as $SW_I = \frac{\gamma/\gamma_{rnd}}{L_{CP}/L_{CP_{rnd}}}$; the more the index gets high the more the network gets close to a small-world. The comparison between effects of line removal shows that the topological measures of characteristic path length increases substantially (it almost doubles) between the situation that do bring to outages and the ones that do not. Therefore the claim of the authors is that static topological analysis can provide an insight about the faults on lines that might lead to outages problems.

Han \etal~\cite{Han2011} perform a \sw analysis of two samples of the Chinese network. The \sw analysis applied to the Anhui Grid does not satisfy the clustering coefficient condition which is slightly higher than the one of a random graph with same \textit{order} and \textit{size}. On the other hand the other sample analyzed representing East China \G completely satisfies the \sw conditions.

Ding \etal~\cite{ding06} in addition to compute the usual metrics to evaluate the satisfaction of the \sw property for the two Chinese \G under investigation, concentrate their efforts on modifying and adapting the \sw model to consider a weighted graph. They adapt the clustering coefficient by weighting the traditional value of clustering coefficient by the ratio between the average impedence of the line in the neighborhood of the considered node and the average impedence of all the lines in the network. Instead of considering the \CPL the authors use a similar quantity that can be applied also to a weighted condition which is the characteristic node number (the median of the mean of the number of node traversed for all the shortest paths connecting each vertex to all the others). The authors claim that the improved model better captures the real characteristics of the \G and it shows more than the unweighted model the compliance to the \sw model in both the two samples analyzed.

Sun~\cite{sun05} evaluates the \sw properties in American, North and Center China \PG and the results show that all these samples can be considered examples of \sw networks. The conditions of \sw are satisfied for the three samples considered, but the Center China \PG actually is a borderline condition especially for the clustering coefficient which is just 4 times higher than a random graph with same \textit{order} and \textit{size}. The author stresses that although the \PG are systems that are developed in a careful manner and each step of development is planned in detail, still  some characteristics of randomness remain which might be due to the many practical factors that involve the \PG plan. The author also remarks a known property of \sw networks that is the ability in easily spread information in the same network. This aspect in the \PG to a certain extent is not always beneficial since this also eases the spread of cascading failures that lead to \PG blackouts.

Pepyne~\cite{pepyne07} uses a \sw network model to evaluate the cascading effects of networks with different values of the rewiring probability parameter. The author also investigates the \sw properties of the IEEE 57 and 118-Bus by verifying the conditions for these type of network defined by Watts and Strogatz~\cite{Watts98,Watts03}; the two IEEE samples satisfy such conditions.

\vspace*{1cm}
\begin{center}
\framebox[0.75\linewidth]
{
\begin{minipage}{0.70\linewidth}
\textbf{Summary:}

In general, the various studies tend not to have a common answer for the general question regarding the membership of \PG networks to the \sw group. It is indeed very specific to the samples analyzed and no conclusion can be drawn, this seems especially true for the \HV Grid, while the \MLV networks seem far from being a \sw network~\cite{PaganiAielloTSG2011}.

\end{minipage}
}
\end{center}

\subsection{Node degree distribution}

The degree of a node is a property to understand how many other nodes it is connected to. However, this information is not particularly important for big graphs since keeping track of each node degree may not be manageable, instead it is better to have a general idea of the statistics of the node degree. In particular, its probability distribution gives us some insights of the general properties of the networks such as the likely or unlikely presence of nodes with very high degree (sometimes also referred as hubs).
Table~\ref{tab:compaND} shows the main information about the degree distribution in the works that perform this study. 
The second column gives a general idea of the type of cumulative node degree distribution that is investigated in the articles under review. What is interesting is to fit the distribution to a class of curves. This is shown in the third column.

\begin{table}[htbp!]

\begin{center}
\begin{footnotesize}
    \begin{tabular}{| c| c |c|}
    \hline
    \textbf{Work}  & \textbf{Cumulative Node  }& \textbf{Fitted }\\ 
            & \textbf{Degree Distribution }& \textbf{Distribution   }\\
      & \textbf{Probability Type }&\\\hline
    \cite{alb:str04}  & \Exp &$y(x)\sim e^{-0.5x}$\\ 
     \hline
   \cite{Crucitti04} & \Exp &$y(x)=2.5e^{-0.55x}$\\ 
    \hline
    \cite{Chassin05}  & \Pl & $y_1(x)=0.84x^{-3.04}$  \\ 
     & &$y_2(x)=0.85x^{-3.09}$\\ \hline
    \cite{Rosas-Casals2009} & \Exp &$y_1(x)\sim e^{-0.81x}$ \\ 
     &&$y_2(x)\sim e^{-0.54x}$\\ \hline
    \cite{casals07}  &\Exp & $y(x)\sim e^{-0.56x}$\\ 
    & &\\ \hline
    \cite{Corominas-murtra2008}  & \Exp & $y(x)\sim e^{-0.61x}$\\ 
     \hline
   \cite{Rosato2007}  &\Exp or &  $y_1(x)=e^{-0.18x^2}$\\ 
    &sum of exponential &$y_2(x)=e^{-0.21x^2}+0.18e^{-0.25(x-4)^2}$\\ 
    &terms&$y_3(x)=0.96e^{-0.17x^2}+0.25e^{-0.19(x-3.9)^2}$\\ \hline
   \cite{Wang2010,wang10_2} &Sum of truncated  & $y_1(x) \sim f_1(x)$ \\ 
    &geometricaland irregular & \\ 
    &discrete terms &  $y_2(x)\sim f_2(x)$\\
    && \\ \hline
   \cite{PaganiAielloTSG2011} & \Pl (unweighted) & $y_1(x)\sim x^{-1.49}$ \\
    &and sum of exponential  & $y_2(x)\sim 0.15e^{-21.47x}+0.84e^{-0.49x}$ \\ 
    &terms (weighted)&\\\hline
    \cite{Han2011} & \Exp & $y_1(x)\sim e^{-0.65x}$ \\
    &  & $y_2(x)\sim e^{-0.58x}$ \\ \hline
    \cite{sun05}  &\Exp & $y(x)\sim e^{-0.5x}$\\ 
    & &\\ \hline
    
    \end{tabular}
     
    \end{footnotesize}
\end{center}

\caption{Comparison of the node degree cumulative
  distribution probability functions. \label{tab:compaND}}
\end{table}

\begin{figure}[h!]
 \centering
\includegraphics[width=0.7\textwidth]{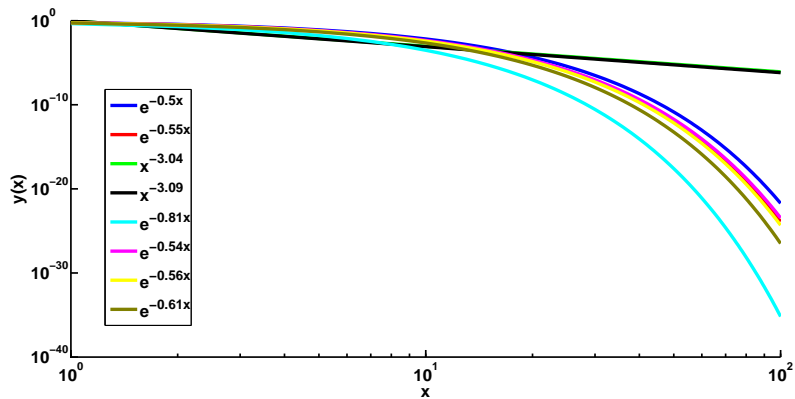}
 \caption{Log-log plot of fitted node degree cumulative probability distribution corresponding to the first six rows of Table~\ref{tab:compaND}.}
 \label{fig:ndd1}
\end{figure}

\begin{figure}[h!]
 \centering
\includegraphics[width=0.7\textwidth]{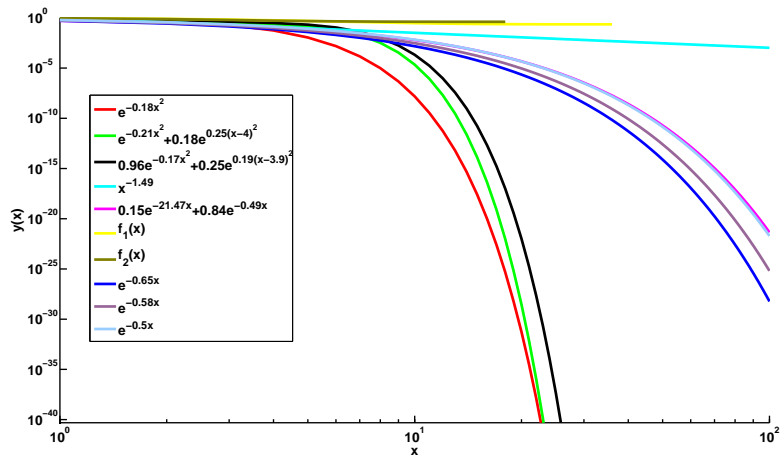}
 \caption{Log-log plot of fitted node degree cumulative probability distribution corresponding to the last five rows of Table~\ref{tab:compaND}.}
 \label{fig:ndd2}
\end{figure}

As seen in the table, the results do not completely agree on the type
of the distribution followed by the \PG networks, but generally they
are close to an exponential decay. Figures~\ref{fig:ndd1} and~\ref{fig:ndd2}
represent the fitted node degree cumulative distribution reported in
the third column of Table~\ref{tab:compaND}. For~\cite{Wang2010,wang10_2}
presented in the table the functions $f_1(x)$ and $f_2(x)$ are not
reported in the table for size reason, but in footnote.\footnote{
    $f_1(x)=\sum_{x_i<x}
    {0.2269(0.7731)^{x_i} \ast \{ 0.4875\delta(1),0.2700\delta(2),0.2425\delta(3)}\}
   $\hspace{3mm}$x_i=1,2,\ldots 34$\\
   $f_2(x)=\sum_{x_i<x}
    {0.4085(0.5916)^{x_i} \ast \{ 0.3545\delta(1),0.4499\delta(2),0.1956\delta(3)}\}$ \hspace{3mm} $x_i=1,2,\ldots 16$\\
    The $\ast$ symbol is here to be considered as the convolution operator and the $\delta$ is the Dirac delta function.}
 The plots in Figures~\ref{fig:ndd1} and~\ref{fig:ndd2} give a general idea of the shape of the distribution. The charts have to be interpreted in a qualitative way since the details concerning the coefficients are not always available in the reviewed studies. In addition, for studies concerning multiple samples (i.e., \cite{Rosas-Casals2009,casals07,PaganiAielloTSG2011}) averages between all samples, or particular significant samples have been chosen among the many available.

The investigation in~\cite{holmgren06} and the charts shown report a node degree distribution for the Western U.S. and for the Nordic \G that both seem to follow an exponential distribution.

The results found in~\cite{hines08} show that the properties of the network (the IEEE 300-Bus blueprint with 300 nodes and 411 edges) change once the weights, in the specific case the impedance characterizing the lines, are considered in the network instead of the traditional topological ones. With impedance as weights on the edges the shortest paths that are computed are actually the ones that the current flow follows according Kirchoff's electrical laws.  In fact, if the graph is rebuilt considering only the 411 node-to-node connections (which are the 411 paths with minimal cost in the original graph representation) with smallest impedance (defined as electrical connectivity) the resulting graph is dominated by few hubs, thus suggesting a scale-free topology when electrical connectivity is considered.


The charts shown in~\cite{wang10_c} that analyze the node degree centrality distribution give an idea about the type of relationship that the node have resembling a power-law. A similar tendency is shown in the chart representing the node degree centrality distribution taking into account the electrical parameters of the network (impedence of transmission lines).

\vspace*{1cm}
\begin{center}
\framebox[0.75\linewidth]
{
\begin{minipage}{0.70\linewidth}
\textbf{Summary:}

In general the various studies focusing on the \HV \Gs agree on a statistical distribution for node degree that follows an exponential (or exponential based) distribution with characteristic parameters of the exponential curve that depend on the specific \G analyzed. While \HV \G have been quite extensively analyzed the \MLV \Gs have not found much attention so far and a deeper and wider investigation needs to be performed in different geographies since the only study (i.e.,~\cite{PaganiAielloTSG2011}) is representative of the Northern part of the Netherlands. In addition, the distribution \G will be the section of the \PG mostly impacted by Smart Grid technology~\cite{brown08}.

\end{minipage}
}
\end{center}

\subsection{Betweenness distribution}

Betweenness is an important measure to assess how a node is central in a network. This metric in fact computes how many shortest paths traverse a node, therefore giving an information of the importance of the node in the path management. The main characteristics of the betweenness study are summarized in Table~\ref{tab:compaBet} where the second column shows the type of followed distribution, while the analytical function is represented in the third column. Unfortunately, this metric is computed by only five studies (~\cite{alb:str04,Crucitti04,PaganiAielloTSG2011,Han2011,wang10_c}).
\begin{table}[htbp!]
\begin{center}
\begin{footnotesize}
    \begin{tabular}{| c| c |c|}
    \hline
    \textbf{Work } & \textbf{Cumulative Betweenness }& \textbf{Fitted }\\ 
            & \textbf{Distribution} & \textbf{Distribution   }\\
      & \textbf{Probability Type} &\\\hline
    \cite{alb:str04}  & \Pl &$y(x)\sim (2500+x)^{-0.7}$\\ 
     \hline
    \cite{Crucitti04} & \Pl &$y(x)\sim 10000(785+x)^{-1.44}$\\ 
     \hline
    \cite{PaganiAielloTSG2011} & \Pl and & $y_1(x)\sim x^{-1.18}$ \\
    & exponential & $y_2(x)\sim 0.68e^{-6.8\cdot 10^{-4}x}$ \\ \hline
    \cite{Han2011} & \Pl &$y(x)\sim x_1^{-1.71}$\\
    && $y(x)\sim x_2^{-1.48}$ \\ \hline
    
    \end{tabular}
    \end{footnotesize}
\end{center}
\caption{Comparison of the betweenness cumulative distribution probability functions.\label{tab:compaBet}}
\end{table}

\begin{figure}[htbp]
 \centering
\includegraphics[width=0.7\textwidth]{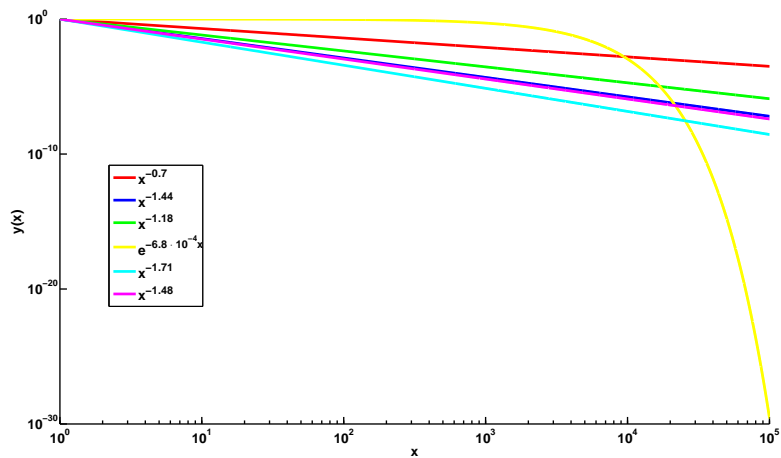}
 \caption{Log-log plot of fitted betweenness cumulative probability distribution corresponding to Table~\ref{tab:compaBet}.}
 \label{fig:betcpd}
\end{figure}

Although the studies that perform this type of analysis are only few, one can see that there is a  tendency for the \HV network to have a betweenness distribution close to a Power-law. For the \MLV the situation is less clear: some samples analyzed in \cite{PaganiAielloTSG2011} follow an exponential decay, especially the Low Voltage ones, while other, usually the bigger belonging to the Medium Voltage, follow a Power-law. In Figure~\ref{fig:betcpd} the plot of the distributions is represented to show the difference between the trend of the Power-law and exponential decay: after a certain point the exponential distribution has a faster decay.

The work of Hines \etal~\cite{hines08} reinforces the idea that betweenness has a \Pl trend in its probability distribution. In particular, the author analyzes betweenness in a condition where weights representing lines impedance are considered. The results show, even without providing an analytical representation, a tendency  for this metric of being dominated by few nodes that support much of the weighted paths, while the majority of the nodes is only slightly involved. This suggests a \Pl distribution.

Wang \etal~\cite{wang10_c} investigate several centrality measures (i.e., degree centrality, eigenvector centrality, closeness, betweenness both on vertexes and edges). The normalized vertex betweenness whose chart is given in~\cite{wang10_c} suggests the presence of a \Pl although the analytical description of the best fitting curve is not given. The authors interestingly note that there is a very good accordance in this metric between the values computed with the electrical parameters (weighted analysis based on admittance matrix) and purely topological one; there is in fact an 80\% overlap in the first 10 nodes most with highest betweenness.

\vspace*{1cm}
\begin{center}
\framebox[0.75\linewidth]
{
\begin{minipage}{0.70\linewidth}
\textbf{Summary:}

\Pl seems the dominating rule for betweenness probability distribution even if few studies consider this statistical property of graphs. To draw a general definitive conclusion regarding this property for the \PG more studies are required. Another aspect to be considered which has not received much attention so far is the study of betweenness statistics in weighted \G models or when power flows are considered instead of the pure topological analysis.

\end{minipage}
}
\end{center}

\subsection{Resilience analysis}

The characterization of resilience is the main motivation for the studies involving \CNA and Power Grid. In fact, the behavior in terms of connectivity of the network when nodes or edges are removed is the primary question in many works considering failures that happen in a random fashion or following an attack strategy. Table~\ref{tab:compaResAna} describes the different types of resilience analysis that are performed by the various authors. In particular, the second column contains the  metric that is used to assess the reliability of the network. The focus of the attack either it is related to nodes or edges is considered in third and fourth column, while the fifth column remarks the studies that in addition to the resilience analysis also propose a mitigation strategy for improving \G reliability.

\begin{table}[htbp!]
\begin{center}
\begin{footnotesize}
    \begin{tabular}{| p{1.8cm}| p{5cm} |p{1cm}|p{1cm}|p{1.6cm}|}
    \hline
    \centering
   \textbf{ Work  }& \textbf{Resilience Analysis Type }&\textbf{Node attack }&\textbf{Edge attack }&\textbf{Grid improvement }\\ \hline
    \cite{alb:str04}  & Connectivity loss &\checkmark & &\\  \hline
    \cite{Crucitti04} & Efficiency&\checkmark & & \\ \hline
    \cite{Chassin05}  & Loss of load probability &\checkmark &\checkmark&\\  \hline
    \cite{holmgren06}  & Influence on largest component size and path length &\checkmark &&\checkmark\\ \hline
\cite{Rosas-Casals2009}& Robustness through mean degree, motifs and patch size analysis&&&\\ \hline    
    \cite{casals07} & Influence on largest component size &\checkmark & & \\ \hline
    \cite{Corominas-murtra2008} &  Influence on largest component size e comparison with theoretical results &\checkmark & &\\  \hline
    \cite{Crucitti2005}  & Damages and improvements&&\checkmark &\checkmark\\ \hline
    \cite{Rosato2007} & Nodes disconnection and improvements&&\checkmark &\checkmark \\ \hline    
    \cite{PaganiAielloTSG2011} & Influence on largest component size &\checkmark &\checkmark &\\ \hline
    \cite{Mei11} & Several criticality analysis and blackout models &\checkmark &\checkmark &\\ \hline
    \cite{casals10} & Reliability and disturbances & & &\\ \hline
    \cite{bompard10} & Unserved energy/load &\checkmark & \checkmark &\\ \hline
    \cite{kim07} & Critical Path Length & & \checkmark &\\
    &and clustering coeff.&& &\\ \hline
    \cite{hines08} & Sensitivity & & &\\ \hline
    \cite{wang10_2} & Influence on largest spanning cluster size & \checkmark & &\\ \hline
     \cite{Han2011} & Loss of load and failure endurance &\checkmark & &\\ \hline
     \cite{Wang2009a} & Avalanche size &\checkmark & &\\  \hline
     \cite{dwivendi10} & Flow availability & & \checkmark &\\ \hline
    \cite{dwivendi09} & Efficiency & & \checkmark &\\ \hline
    \cite{guohua08}& Largest power supply region &\checkmark &\checkmark &\\ \hline
     \cite{pahwa10}& Influence in network connectivity & &\checkmark &\\ 
     &and power degradation&&&\\ \hline
     \cite{kinney05}  & Efficiency &\checkmark & &\\  \hline
     \cite{arianos09}  & Efficiency, net-ability, overload & &\checkmark &\\  \hline
     \cite{bompard09}  & Efficiency, net-ability, overload & &\checkmark &\\  \hline
     \cite{hines10}  & Path length, connectivity loss, &\checkmark && \\ \hline
     \cite{pepyne07} & Line overload, cascade effects,& &\checkmark  &\checkmark\\
     &network disruption&& &\\\hline
     \cite{Brummitt12}  & Overload, cascade effects, &\checkmark & &\checkmark\\ 
     &blackout size&& &\\\hline
    
    \end{tabular}
    \end{footnotesize}
\end{center}

\caption{Comparison of the resilience analysis and improvement studies.\label{tab:compaResAna}}
\end{table}

\subsubsection*{Node based attack analysis}

Albert \etal~\cite{alb:str04} show the cascading effect of the whole American \PG when removing a certain fraction of nodes. In particular, the authors define the concept of \emph{connectivity loss} which expresses the magnitude of the substations that cannot receive power from any generator due to failures in the network and thus inhibiting the end users as well to receive any power. The connectivity loss assumes different levels of severity based on the type of nodes, i.e., transmission substations, that are removed. The random removal of these nodes has a limited impact on the connectivity loss, that increases almost linearly with the removal; the situation is completely different if the removal targets the nodes with highest betweenness especially when the betweenness is re-calculated after each new removal. In this situation, the connectivity loss increases harshly after a certain number of substations are compromised and just removing 2\% of the nodes brings to a connectivity loss of about 60\%; the behavior is the typical non-linear one that characterizes threshold effects.

Crucitti \etal~\cite{Crucitti04} propose a model that considers \emph{efficiency}, a measure inversely proportional to the shortest path. In particular for each node the authors define weights both for nodes (i.e., the maximum capacity a substation can handle, taking also into account a certain level of permitted tolerance or overcapacity) and for edges (i.e., the ability of delivering power for a certain transmission line). To study cascading effects after the removal of one node, the edge weights are re-calculated to investigate what effects on the system are triggered, and what is the new ability of edges in supporting paths. They show the cascading effects and the problems the network suffers in term of \emph{efficiency} in different situations: random node removal and highest load-based removal. The results show a dissimilar behavior in the two situations especially when the tolerance parameter is low (i.e., the substations are considered to operate with small possibility to sustain more than the nominal capacity). Also in this case the worst results are experienced when the nodes with highest load are removed one after each other. Perhaps surprisingly, there is a non-perfect correlation between node degree and betweenness: it is not implied that the nodes with highest degree have always a high load.

Casals \etal~\cite{casals07} investigate the behavior of the European \G under failures or attacks. The analysis shows a typical pattern already known for many other type of networks: the giant component (i.e., the biggest connected component) of the network vanishes after a threshold of nodes is removed and a phase transition occurs. The theoretical threshold that is computed for exponential node degree-based graphs well suits the empirical details that are extracted from the samples. Under random failures the decrease in the {\em order} of the giant component is continuous until the threshold is reached; under targeted attacks (e.g., removal of nodes with highest degree) the samples show a network disruption that appears when a smaller fraction of nodes is removed. A remarkable result the authors find is the relation between the {\em order} of the network and the {\em order} of the giant component: an increase in the number of nodes of the network makes it more prone to failures, but at the same time the way the networks have evolved tend to reduce their fragility.

Sole \etal~\cite{Corominas-murtra2008} investigate the consequences of intentional attacks European \PG might face. The targeted attack problem is translated into an equivalent problem of random failures such that it can be studied with percolation theory, thus identifying a threshold of nodes to be removed to breakdown the giant component of the network. The fraction of nodes to be removed in targeted attacks towards highly connected vertexes is, as intuition might suggest, much smaller than in random attacks. The results show that there is generally an acceptable matching between the theoretical parameters found and the value extracted from the samples, nevertheless some deviations are present especially for those networks that are more robust. A remarkable aspect is that there is a correlation between the critical fraction removal and the most important reliability indexes used by power engineers to measure the \PG performances (e.g., energy not supplied, loss of power, interruption time). In fact, the two groups of European Grids that are considered (based on the similarity of the calculated and theoretical value of the threshold) although managing almost the same amount of power and energy, show very different results in terms of failures. The networks with threshold that deviates positively from the theoretical values are much robust and experience small reliability issues, thus reinforcing the correlation between non-topological reliability indicators and the topological ones.

Wang \etal~extend forward in the analysis done in~\cite{Wang2010} and in~\cite{wang10_2} by studying the vulnerability properties of four synthetic IEEE Bus models and two real networks (NYISO and WSCC). They  evaluate the robustness of the samples in two distinct situations: random node removal and targeted attack towards node with highest node degree. They evaluate the presence of a threshold value in the number of nodes removed before the network spanning cluster (i.e., a notion equivalent to the largest connected component of a graph) is fragmented. The authors reach the non-surprising conclusion that networks are more vulnerable if the nodes with the highest degree are attacked first compared to a random attack. The main finding is the definition of the equation that computes the critical fraction of nodes that destroy the spanning cluster in targeted attack conditions. In fact, they find that this value is influenced by the node degree distribution of the network. As a better fitting to the theoretical value of the critical fraction (threshold) the authors propose to consider the node degree distribution is modeled with a truncated geometric distribution plus an irregular discrete distribution~\cite{Wang2010}, rather than an exponential one as computed for the critical fraction in~\cite{Corominas-murtra2008,casals07}.

Han \etal~\cite{Han2011} consider two elements to characterize  problems in the Power Grid: load operating limit and the maximal load that a node is able to support. The reactance value of lines is adjusted accordingly when the operating limit is overtaken and the line is considered discarded if the load of the node to which it is connected is higher than the maximum supported value. The mechanism for evaluating the spread of failures in the network is by rerouting the paths when the nodes (and the edges attached to them) are not available. To show the impact of failures the authors use a loss of load metric which is given by the ratio between the load of nodes which are no more in operation (since they have overtaken their limits) and the initial load supported by all the nodes. This loss of load metric is applied to evaluate the effects of two sorts of failure modes: nodes with high degree and high load are removed first and the effects are evaluated; nodes are divided in groups and average load for each group is computed, then the first 20 groups of nodes in terms of their load are removed, a second mode of evaluation is by random group removal and after each removal an increase in capacity factor of lines by 0.0125 is performed. These removal methods are applied to both the Anhui and East China \PGs and they show for both networks that the  spread of failures when attacking high degree nodes spreads quite slowly; on the other hand, when the nodes with high load are attacked the \sw network (i.e., East China Power Grid) appears particularly vulnerable and causes an important decrease in transmission capacity. The second type of attacks (based on group of nodes) evaluates the effects on the reduction of load by a group of nodes removal compared to the effects of the removal of the first group of nodes; this is to test the endurance capability of the network. The results show that Anhui \G (non small-world) improves its performance thanks to the increase in capacity at each round, on the other hand the improvement is extremely limited in the East China Power Grid. The authors claim that the weaknesses of the \sw network compared to the other under test is due to the very nature of the network itself because of the high heterogeneity the nodes have in their load, with few nodes (those that are attached to the ``shortcut'' edges) having to sustain heavy load.

Wang \etal~\cite{Wang2009a} consider the behavior of the network and the cascading failures issues when nodes with high and low load are attacked respectively. Conversely from other works the definition the authors provide in their work for load is obtained by a product of the node degree and the sum of the node degree of the neighbors of the same node amplified or lessen by a tunable parameter that represent the initial load of a node. In addition each node is characterized by a tolerance parameter which directly influence the capacity (maximal load) that the node is able to sustain. If the limit is overtaken than the node is considered broken and the load sustained is redistributed to its neighboring nodes proportionally to their initial load. A node that is under attack has to redistribute its load to its neighbors. The metric used in to understand the problem induced by an attack are measured through an avalanche size indicator that considers how many nodes reach the failure condition after an attack is performed. The authors consider two strategy for attacks: the first strategy involves nodes with highest load that are attacked in descending order and the second strategy attacks nodes starting from those with lowest load in ascending order. The authors find that, based on the initial load that is given to nodes, it is more likely that one strategy or the other is more critical in triggering avalanche failures. In particular, when the initial load is limited the cascade effects are triggered by the nodes with the lowest one, the situation is reversed when the initial load is bigger: highest load nodes attacks generate cascade effects behaviors. Typical of this work is the presence of a threshold in the tolerance of the lines: if the tolerance is above the threshold then the network do not experience avalanche effects, but if the threshold is even slightly overtaken the cascade effect arises. 

Kinney \etal~\cite{kinney05} perform a comprehensive analysis of the failure behavior of the U.S. Power Grid. The metric that is assessed is the damage that is inflicted to the network computed as the loss in average efficiency (defined as the average over all paths of the harmonic composition of the efficiencies of the edges composing the path) of the network before and after the breakdown has occurred. The mechanisms considered for the simulation of the failures are two: 1) single node removal and 2) progression removal of nodes. For each of the two schemes two different methods of node selections are used: a random choice of the node or nodes to remove and a targeted selection involving the nodes that support the highest load. The results found show that in low tolerance conditions (i.e., the capacity of nodes is very close to their initial load) just the removal of a node can cause a damage equal to 25\%. In general the authors find three categories in which the nodes can be fitted: nodes whose removal causes no or very little damage independently of the tolerance parameter, and these are about the 60\% of the overall nodes of the U.S. Power Grid; very critical nodes whose removal causes high damages in the network even when the tolerance is high; nodes whose removal causes damages depending on the tolerance of the network, usually a threshold value exists that is responsible for a steep transition from a little damage to an high damage situation, this last type of nodes are the majority in the network. Usually, nodes that cause critical effects if removed are also characterized by topological properties. In particular, nodes causing no efficiency loss have both low betweenness and low node degree: 90\% of the transmission nodes that cause almost no damage to the network have 2 as node degree and load smaller than 2000. The main finding is once again that the removal of nodes with high betweenness and high node degree are the most critical and just the removal of 0.33\% of nodes (those with such characteristics) generates more than 40\% of damage in the network. The work suggests that changes in topology by adding more transmission lines and substation can help in decrease the sensitivity to failures of high load transmission lines.

Hines \etal~\cite{hines10} have the goal of comparing the vulnerability analysis obtained considering \G models based only on topological properties and \G models that take into account also the electrical/physical properties in order to assess the goodness of topological analysis in \PG vulnerability problems. To perform this comparison three metrics are evaluated, two of them are typical of the \CNA literature and the third is more related to the electrical behavior of the Grid. In particular, the first metric assessed is the \CPL so that as the components in the \G fail the effort in the reachability of other nodes is evaluated compared to the initial situation. The second topological parameter investigated is the connectivity loss (metric already used in~\cite{alb:str04}) which considers the percentage of generating nodes that are unreachable  due to failures in the network. The third metric takes into account the blackout size as the amount of power load that has been curtailed as a consequence of the cascading failures in the Power Grid. In order to evaluate this last metric a model considering physical/electrical parameters and the direct current power flow  is proposed (see description for~\cite{hines10} in Section~\ref{sec:statProp} for the details of the model). The authors consider different types of failures that the network is subject to: random failures of nodes, degree-based attack which removes nodes one after each other starting with nodes with the highest degree. Another attack is based on the maximum traffic that a node manages in terms of power that flows through it and the attack targets the node with highest traffic; the opposite situation is also considered: minimum traffic attack that removes nodes starting with those having smallest traffic. The last attack considered takes into account betweenness and nodes with highest betweenness are removed first. Considering the path length metric the attacks based on degree, max-traffic and betweenness provoke grater damage in terms of path length compared to a random removal of nodes; the min-traffic has a behavior almost the same as random failures. The same considerations can be applied for the connectivity loss metric showing that the \PG is more  vulnerable to directed attacks than random, a behavior thus more similar to scale-free networks. Generally the same considerations apply to the blackout size metric too, but it worths to highlight that just by attacking the 10 nodes with highest max-traffic creates a blackout with average size of 72\%, while a random attack on 10 nodes on average results in a blackout of just 20\% while the removal of 10 nodes with min-traffic results in 5\% average blackout. Although the general intuition shows that attacks that target nodes with certain properties are more damaging than random or min-flow attacks it worths to remark that for the three metrics considered (i.e., path length, connectivity loss and blackout size) there is no accordance on the type of attack that causes the most damages: in the first case it is betweenness, in the second case it is node degree and for blackout metric it is max-traffic that provokes the most damage. The authors conclude that topological measures provide some indication of the vulnerability of a network, but they remark also that if taken alone they could be misleading especially if mitigation strategies on failure reduction are based on one topological result only. The authors also stress that physics-based models are more realistic and generally more useful for infrastructure risk assessment than topological only models.

\subsubsection*{Edge based attack analysis}

Kim \etal~in~\cite{kim07} replicate two blackout conditions that happened in the United States in 1996 by inspecting the topological properties of these networks during the cascade events that brought to the blackout. In particular, the authors simulate on the \PG graph the power line trimming process events that happened and other line removal events that do not lead to outages in the network. The metrics that they inspect to assess the vulnerability are two static topological metrics: clustering coefficient and characteristic path length; they also combine them into the ``small world-ness index'' as described in Section~\ref{sec:sw}.

Dwivedi \etal~\cite{dwivendi10} consider the vulnerability that an IEEE 39-Bus is subject to when the most central lines are removed. The centrality metric developed in this work considers the flow that power lines are able to support given their physical properties (admittance values): the lines that carry more flow (appropriately normalized to make the comparison fair between the power supported by the different source and sink nodes) are the most central. The authors claim that this method of investigating the most critical lines is more precise and accurate than the purely topological one, since it takes into account the amount of power that a certain line provides. However, the authors specify that some of the lines identified as important through this weighted centrality method are the same that can be identified with a pure topological investigation, thus stressing their importance also from a topological point of view. The same authors in~\cite{dwivendi09} investigate the vulnerability of IEEE 39-Bus and IEEE 118-Bus. Even in this case they consider a weighted approach for the network characterized by his reactance parameters as weights. Betweenness values for the weighted networks are computed for each line so that it is available a ranking of the most critical or vulnerable lines. For each IEEE synthetic model two strategy of network disruption are applied: random line removal and removal based on the ranking of betweenness. The results are evaluated considering the efficiency of the network computed for each line as the inverse of the the reactance. The simulations show results which are not completely unexpected and that are common for \PG networks: a quite good robustness towards random attacks (small reduction of network efficiency); the situation changes when the most critical lines are attacked for which there is a considerable reduction in the network efficiency (60\% reduction for the 39-Bus and almost 40\% for the 118-BUS) by just removing the 10 most critical edges.

Pahwa \etal~\cite{pahwa10} consider the effects of cascading effects in synthetic models when lines are removed. In particular in their analysis they categorize the links in two types: vulnerable and non-vulnerable depending on the damage they create in the network (more or less than 10\% upon their removal). The metrics that are computed to assess the damages are the influence in network connectivity which is purely topological and the Power Degradation which is the fall in the total load supported by the network compared to the original supported load. The authors propose synthetic models of networks generated in random fashion with same \textit{order} and \textit{size} of the reference engineering literature (i.e., IEEE, WSCC) samples. The results show that the generated networks achieve a new equilibrium situation after a cascading is initiated in less steps, while the literature samples stabilize at a later stage when nodes or load is removed. 

Arianos \etal~\cite{arianos09} analyze the vulnerability of two samples of the IEEE literature. The authors compute three set of metrics: efficiency already applied in \PG studies~\cite{Crucitti2005,kinney05}, net-ability and overload. The net-ability takes into account the path from a generator node to a load. It is proportional to the power share of the considered path in transmitting power from generator to load and inversely proportional to the {\em{electrical distance}} (i.e., a function to weight the edges based on the path traversed, the power factor and the impedence of the involved lines) between generator and load. It turns out to be proportional to the power transmission distributor factor for the line and the impedence of the same line. An overload metric is computed by considering the sum over the whole network of the ratio between the power flow of lines and their power flow limit. The difference between the overload condition before and after the removal of a line is the assessed parameter. The test that is performed on the network consists of line removal: for each line removed separately the three metrics mentioned above are computed. The main finding of the analysis is that the metric based on efficiency do not show much difference in the reduction of efficiency of the network when different lines are removed from the network, thus it does not provide much information about the criticality of a line. On the other hand, the net-ability and the overload metric present peaks values when specific lines are removed, thus being more precise to localize potential failing lines. Using the same set of metrics just described the critical lines analysis is applied to the a small synthetic Bus (34 nodes) and to the Italian \HV Grid in~\cite{bompard09}. The main idea is to show how different are the results when metrics based on pure topological parameters are applied compared to the results obtained when also the physical characteristics of lines and nodes are also taken into account. The analysis performed on the effects of a single transmission line removal both in the synthetic sample and in the Italian \PG show that the net-ability metric captures better the lines whose fault results in worse disrupting conditions for the network compared to the efficiency metric. The same conclusion i.e., better adequacy of the metrics that take physical quantities into account is obtained also considering the \textit{entropic degree} compared to the traditional node degree to rate importance of nodes. \textit{Entropic degree} is defined considering not only the number of connections that a node has, but also the weights that the lines attached to the node have considering their impedence value~\cite{bompard09}. The conclusion is that pure topological node degree may underestimate the real importance the nodes. It is interesting to remark that the authors of~\cite{arianos09,bompard09} claim that the critical lines identified by using the net-ability metric are the same that the Italian \G operator identifies as critical using engineering and electrotechnical approach to \PG failure analysis.

Pepyne~\cite{pepyne07} evaluates the cascading effects on IEEE blueprint networks. The author uses a DC flow model to characterize the network and to assess the cascading effects: each line is characterized by an impedence and at each node with generating capabilities a power flow injection is randomly assigned as well as the load nodes demand. In the study the line with the highest load is then brought to the  99\% of its trip point and subsequently overloaded to study the evolution of power redistribution process and eventual subsequent failures of other lines. The condition the author evaluates to consider the cascading effects are two: 1) the cascading sequence proceeds till the network breaks apart into disconnected islands; 2) the cascade is self-limiting and stops before breaking up the grid. The simulation on the IEEE 57-Bus sample shows that just after 4 lines experiencing outages, the network breaks apart. The author evaluates also the behavior of \sw networks while increasing the rewiring probability of edges from a situation close to a regular lattice (rewiring probability $p=0.001$) and a situation closer to  a random graph (rewiring probability $p=0.5$).  The author finds that with an increase in rewiring probability (i.e., which corresponds to an increase in the number of so-called ``shortcut'' edges) there are two effects: 1) the load of the network increases and the line congestion decreases (measured as the mean and maximum line loading); 2) the network is more fragile in cascading failures than regular topologies since it tends to break apart just after few lines are in failure state. The author also tests a solution based on the upgrade of lines that in the IEEE Bus model act as kind of ``shortcut'' links whose capacity is doubled. The simulations with improved lines show that the average load of lines reduces thus being able to accommodate more load on average, on the other hand since the maximum loadability of lines is higher, more load has to be redistributed in case of failures, therefore the overall effect is a slightly more fragile Grid. The author finally suggests possible improvements for the real Grids: adding new ``shortcut'' links and increase the capacity of the existing ``shortcut'' links.

\subsubsection*{Node and edge based attack analysis}

Chassin \etal~\cite{Chassin05} define a failure propagation model that considers failure probability for nodes and failure probability for edges. For the latter, although the model is based on an undirected graph, the propagation of failures happens in a directed way, i.e., the propagation evolves when the power flows from the node and the node supports the edge. The two probabilities of failure are then combined together to determine the \emph{Loss Of Load Probability}. The results obtained in term of Loss of Load Probability are similar to the ones obtained for other electrical systems by electrical engineering studies, supporting the validity of the model.

In~\cite{PaganiAielloTSG2011}, we perform a reliability analysis considering different node removal strategies: random, node degree-based, betweenness-based. We evaluate the {\em order} of the largest connected component while nodes are removed. The results obtained show a general robustness of \MLV networks to random attacks while they are extremely vulnerable to targeted attacks. The study also investigates the reliability regarding attacks towards edges exploiting the Laplacian matrix spectrum method to identify the most critical edges.

The book~\cite{Mei11} focuses on evaluating and modeling the reliability of the \PG considered as a system which is characterized by a self-organized criticality behavior and modeling the behavior of the \G that drives to blackout situations. The vulnerability assessments performed are several. To evaluate the reliability of the network the authors consider the static behavior (i.e., in steady state conditions) of the \PG when lines are removed. The authors develop a set of indexes to evaluate 1)~the impact the faults has on active power transmission 2)~the influence of the fault in local balance of reactive power. The first index is called {\em{average electric transmission distance}} and considers the ratio between the weighted transmission path length (a weighted path where the weights are the product of active power transmitted and physical length of the line) and the total power received by nodes; this index is a global measure that describes the influence of network structure on transmission of active power. The second pair of indexes takes into account the local balance in reactive power in particular the first index considers the difference in the voltage experienced before and after the fault, the second considers the variation in the reactive power supplied by generators before and after the fault compared to the maximal reactive power they can provide. The global and local metrics are then combined into a single vulnerability index. Using this combined index the authors then assess the vulnerability of the transmission lines of IEEE-30 and IEEE-118 Bus systems and a snapshot of Shanghai \PG of 2006. They also compare the results obtained through the method proposed in traditional power electrical analysis related to lines criticality;  the results show a good agreement in identifying the most vital lines for the operation of the Grids. In another section of the book an investigation of vulnerability more in line with traditional high level Complex Network evaluates the behavior of \PG generated following the \sw model proposed by the authors for \PG evolution. They consider several types of attacks on nodes (random node, static node degree, dynamic node degree which is recalculated after each removal, node betweenness) and on lines (random line, line betweenness). The metrics they evaluate are similar to the ones of other works such as efficiency in transmission calculated as the ratio between the sum of the reciprocal of the shortest path between any two nodes after the failure has occurred and before the problem has taken place. All these types of attack and the metrics computed are applied to \sw generated networks, a 300 Bus network, IEEE-30 and IEEE-118 Bus systems. The main distinction between the \sw and IEEE networks lies in a better reliability of \sw networks towards random failures compared to IEEE, on the other hand IEEE Bus systems report better scores when the attacks are targeted (i.e., betweenness of degree based). The book also models the blackout situations occurring in a power systems considering a model based on DC (direct current) power flow and AC (alternating current) from a more electrical (power flow equations and power system constraints are applied) than topological point of view. These models are respectively applied to the \HV \PG of North East China and to IEEE-30 Bus system.

Bompard \etal~\cite{bompard10} considers betweenness as the metric to investigate in analyzing the vulnerability of Power Grids. Unlike the other studies~\cite{alb:str04,Crucitti04,PaganiAielloTSG2011,Mei11} which consider betweenness only from a topological point of view, the author builds a graph which has on each edge a weight that represents the value of power flow that is managed by the line; the betweenness computed with such a graph is called electrical betweenness. The metric assessed to evaluate the removal of nodes, or edges, with highest electrical betweenness is the {\em{unserved energy}}. The results  obtained by removal of edges, or nodes, considering the electrical betweenness are more critical compared to the ones found in computing topological betweenness since the quantity of unserved load is higher in the first case. The author therefore claims that electrical betweenness is a more correct indicator than topological betweenness when assessing vulnerability of Power Grids.

Guohua \etal~\cite{guohua08} evaluate the vulnerability of the  North China Power Grid. As the metric to evaluate the effects of the disruption the authors propose the largest power supply region index which is obtained by the ratio between the maximum load that the system can handle after a cascading failure has occurred, and the initial total load of the system. To assess this metric different disruption policies are applied. Random failures happening on lines are simulated and the probability law governing the failures of the lines is a Poisson distribution. A second type of attack involves nodes and they are attacked in descending order of node degree. Third type of attack considered deals with nodes that have the highest vulnerability index: in the weighted and directed model of network the authors consider, each node has a supported input power that gives it a certain weight, but also the lines to which it is attached (that have a transmission power limit which is the line weight) are an important part to define the node vulnerability index which turns out to be the product of the node weight and the  line weight. The last policy considered in the paper is a targeted attack on lines based on their descending line vulnerability index which is equal to the betweenness value for an edge when the graph is considered weighted and directed. The results on the North China \PG following these attacks show that the \G supports quite well random failures of lines since no cascading effects happen. On the other hand, the selective attack on high degree nodes are more damaging increasing the spread of damages in the network. Even more critical for the \G are the attacks  based on node vulnerability index and line vulnerability index which generates serious disruption in the network by leading to cascading effects.

\subsubsection*{Other type of robustness analysis}

Casals \etal~\cite{Rosas-Casals2009} find that each single \G composing the European network has a characteristic parameter that is related to its specific \emph{robustness}. The authors investigate three main elements to assess the robustness of the different Grids: mean degree distribution, motifs analysis and patch size distribution. The first indicator is related to the comparison of the average degree of each Grid sample with the average degree obtained by the best fitting Poisson distribution (Poisson distribution for node degree arises from a random graph). The assumption is that when the number of nodes is sufficiently big the exponential tail and the Poisson one are very similar; the authors suggest that, the more the average node degree deviates from the Poisson distribution value, the most fragile is the network. This might seem counter-intuitive, but the networks that deviate more from the random topology are more fragile. The second idea is the investigation of motifs (i.e., linear, stars, triangles) inside the network. The conclusion the authors draw out is that fragility increases as the elements of the Grid become more interconnected and motifs such as stars and triangles begin to appear. Patch size distribution is the third element taken into account to characterize the relation between topology and robustness of network. The results suggest that a balanced distribution i.e., having a constant frequency of patches with different areas, characterizes more robust networks. The ideas and results presented are original and interesting even if, as the authors explicitly mention, a more deep analysis and comparison with Grid's dynamical results are needed.

Rosas-Casals~\cite{casals10} performs a slightly different type of study to evaluate the statistics of the European \HV Grid. It evaluates the statistical distribution of three metrics that are common in the assessment of the performance of electrical systems adopted by electrical engineers: energy not supplied, total loss of power and restoration time. Information about the statistics of failures are obtained by the ENTSO-E (European Network of Transmission System Operators for Electricity).  Based on the results of his previous study of European \G classification as robust or fragile networks~\cite{Rosas-Casals2009}, the author considers the statistical distribution of the three reliability measures investigating the presence of \Pl characteristics in their distributions. The presence of \Pl distribution is plausible only for the restoration time statistics. One of the interesting findings is that robust grids accumulate less failure events than the fragile ones, but the events that strike robust networks lead usually to important consequences, on the other hand fragile networks have more frequent problems but their consequences are usually limited.

Hines \etal~\cite{hines08} consider a different metric in assessing the vulnerability of the network that is based on the physical properties (line resistance) of the network: sensitivity. Sensitivity, which is defined as a ratio between a resistance of a path from one node to any other node and the resistance of the node itself, expresses the extent to which a state change in one node propagates to another node. 
The results show that sensitivity decades linearly with the distance between nodes, but the more the network has a scale-free structure the more the number of nodes are involved. For the IEEE 300-Bus the results suggest a behavior in the decrease in sensitivity with distance between nodes similar to the one of a random graph with same number of nodes and edges.

The vulnerability analysis performed by Sun~\cite{sun05} is actually a description of results obtained by other studies (e.g., \cite{Crucitti2004a,kinney05,Lai04,motter02}) on Complex Networks and \PG studies on the behavior of networks under failures of nodes or edges.

\subsubsection*{Studies with infrastructure improvement analysis}
Holmgren~\cite{holmgren06} inspects the structural vulnerabilities of \PGs considering the decrease in size of the largest component of the graph. In particular, the author compares the resilience to failures of the Nordic Grid, the Western U.S. Grid, a random graph and a Barab\'{a}si-Albert scale-free network. The comparison is considered both with random node removal and with targeted removal focusing towards nodes with highest node degree and proceeding in decreasing order; the size of the largest component is the evaluated parameter. The simulation shows that removing nodes following descending node degree, especially recalculating after each removal, brings the highest disruption in the network. In particular, the Nordel network seems less robust than U.S. Grid that in turn performs close to a scale-free network for these targeted attacks. The author also shows the evolution of the average path length when nodes are randomly removed, highlighting a similar behavior of the U.S. and Nordic Grids, both dissimilar from the behavior of the Barab\'{a}si-Albert and random graph models. The author then proposes a virtual scenario of a very simple Grid for which he analyzes different improvement strategies and different types of potential attacks (both natural circumstances and human-driven sabotages) involving different type of nodes and edges. The author also shows the decrease in vulnerability that different strategies of network improvement may bring.

Crucitti \etal~in~\cite{Crucitti2005} aim to detect the most critical lines for Italian, French and Spanish \HV \Gs and to propose solutions to possible vulnerabilities. They use  \emph{efficiency} as main metric to evaluate the performance of each Grid, that is, a function of the shortest path connecting two nodes. Based on this metric the vulnerability of the Grid is measured in terms of efficiency loss experienced after a damage is inflicted to the network. By computing metrics on experimental data, the authors find significant differences between the Grids. The Italian Grid is the most vulnerable: by removing just one edge (of course the most critical) the decrease in efficiency is of 5\%, while for the Spanish and the French ones the reduction is only of 3\%.  With this approach it is possible to go forward and identify the most critical n-edges that if removed simultaneously decrease efficiency. The worst case occurs when three lines are removed together as experienced by the Italian Grid whose efficiency decreases by 31\%. In addition, the resulting Grid is then broken into two subgraphs. The authors also consider the improvements that might be brought by the addition of one new line to the Grid, the Italian Grid is the one getting the greatest benefits (8\% in efficiency increase). The paper also shows the results of the probability distribution of causing a certain damage (decrease in relative efficiency) for the Grids. The differences in the shape of the distribution between French, Spanish and the Italian are significant: French and Spanish have a Gaussian shape, while the Italian damage distribution is bimodal (obtained by the sum of two Gaussian with different means), in particular, the second peak corresponds to the situations in which the removal of edges brings to the complete break of the network in two subgraphs. 

Rosato \etal~\cite{Rosato2007} use the spectral properties of the Laplacian matrix considering the second smallest eigenvalue and corresponding eigenvector to identify edges that are extremely critical. For the Italian \G the authors find the edges, only three, that if removed break the network apart, while for French and Spanish Grids the value is seven. This result is in line with the findings described in~\cite{Crucitti2005}. The authors also compute the conditional probability of having a certain number of disconnected nodes from the \G once a prefixed number of edges is removed at random. The resulting measure is a sort of robustness index that indicates the level of vulnerability of the network. The results show the most critical is the Italian \G for which the removal of just two edges leads to the probability of disconnecting more than two nodes that is around 5.8\%, while for instance the same situation for the French \G has only a probability of 0.3\%. This is due to the specific morphology of Italy that spans across a long peninsula. A solution to improve the vulnerabilities of the Italian Grid is to add an edge between two strategic nodes, then the number of edges to be removed to break apart the network in two components raises to twelve. In this work, unlike~\cite{Corominas-murtra2008}, the geographical structure seems to play an important role and it is a key to understand certain properties.

Brummitt \etal~\cite{Brummitt12} analyze the cascading effect happening on a \PG sample of the South-East of the United States using a sand pile model to generate cascades. In synthesis sand pile model assumes that each node is characterized by a sand threshold it can support (e.g., maximum load that it can support) and at each time step a grain of sand is randomly dropped on a node; this added sand can trigger the overtaking of the threshold for a node which is considered in a failure state and its sand is redistributed among its neighbors. This redistributed sand (e.g., load) can then trigger other failures in a cascade effect. For more details on the sand pile model refer to~\cite{bak87,bak88}. The analysis in~\cite{Brummitt12} is centered on the effects that connectivity has in networks. In particular, the authors focus both on randomly generated network, with similar node degree distribution to real \PGs and on two real Power Grid, samples. The peculiarity of the samples, which is imposed also in the synthetically generated networks, is the loose connectivity between the two Grids: the two samples of the US \G are actually connected by only 8 edges. The authors find that adding connectivity between the two loosely connected samples is not always beneficial. In particular, an optimal connectivity exists that minimizes the chance of cascading spread. Above this threshold of connectivity, cascades, especially of large dimensions, are likely to spread across the entire connected network. The situation is even more complex when the networks loosely connected have a disparity in the load they support: for one network more external connectivity would be beneficial mitigating his internal cascades, while for the other the increased connectivity would pose more threats. This situation creates the necessity for one \G to increase its capacity in order to diminish the possibility of inflicted cascades originating in the neighboring Grid, thus producing a situation of disequilibrium difficult to solve (e.g., each \G might aim at achieving his local optimal situation without considering the effects on the neighbor).

\vspace*{1cm}
\begin{center}
\framebox[0.75\linewidth]
{
\begin{minipage}{0.70\linewidth}
\textbf{Summary:}

In general, the reliability is assessed by evaluating the connectivity or the ability to efficiently guarantee paths between nodes when nodes or edges in the network are removed. As a general result for failures related to nodes all the samples show a good resilience to random breakdowns. In fact, the network is always able to guarantee a certain connectivity until the number of nodes removed are the biggest part of it. On the other hand the Grids are extremely vulnerable to targeted attacks, that is failures that focus on key nodes for the entire network such as high degree nodes or nodes with high betweenness or nodes or lines that manage the highest amount of load or electricity flow. The thoroughness and space given to the vulnerability and blackout topics in~\cite{Mei11} covering also aspects more related to the electrical domain than pure Complex Network Analysis, provide a view of the reliability of the power system that is more precise and complete.  This last aspect (i.e., using also physical/electrical parameters in the analysis) is stressed also in other works~\cite{arianos09,bompard09,hines10,Mei11,hines08} which provide motivations and results justifying the superiority of vulnerability analyses that have a more realistic view of the \PG network. Adding physical parameters to the network is beneficial to have results concerning the way networks tend to disrupt and spread failures closer to reality. However, an aspect that is missing is the cross-check to real problems experienced by the transmission operators on the lines or nodes of the network identified as most critical and the resulting cascading effects obtained by a pure topological analysis or a topological analysis enriched with physical parameters and constraints characterizing the lines and nodes.

\end{minipage}
}
\end{center}

\subsection{Further studies}
Some of the studies considered in this survey provide additional analyses and investigations that are not common to any other study and thus deserve special attention. In this subsection we describe the most interesting additional aspects of some works for the sake of providing a more complete and thorough picture of \CNA approach to the Power Grid.

Mei \etal~\cite{Mei11} consider the whole electrical system as a Complex System: a system that is characterized by a self-organized criticality. The study does not only deal with a pure \CNA study (as many of the other here surveyed), but it provides more insights into the peculiarity of the electrical \G and systems. In particular, the book goes further in analyzing and modeling the blackout with particular regard to the dynamical aspects of the power system such as the power flow distribution and load redistribution. In addition to this enhanced perspective the book provides a section dedicated to modeling the growth and evolution process of the \HV \PG which is a topic which has deserved small attention so far by \CNA studies.

In his statistical analysis of reliability of Europeans \Gs Casals~\cite{casals10} goes deeper in analyzing the causes reported by transmission operators for the failures experienced. The data shows that only about 5\% of the failures are caused by overload of the lines, while about 30\% of the failures is classified as ``unknown'' (the other causes of problems are categorized as external and failures which correspond respectively to almost 30\% and almost 40\%). As remarked by the author this last  cause of accidents need to be better understood and categorized in order to have a better understanding of the phenomena that trigger failures in the Power Grid.

Wang \etal~\cite{Wang2010} take into account also the electrical properties of the network considering the admittance matrix expressed as a combination of network adjacency matrix and line impedances. The finding of this study on line impedances shows for this property a probability distribution with heavy tails (i.e., some long lines resulting in very high impedance) that is best fitted to a double Pareto lognormal distribution that experiences an exponential cut-off in the tail part. This is due to economical aspects of construction and maintenance of long lines. The authors define a model to generate power networks that is based on creating small size sub-networks, then connect them together in a sort of lattice topology and finally generate the impedances following a certain distribution (e.g., double Pareto lognormal).

In~\cite{PaganiAielloTSG2011}, we point out an important difference resulting in the comparison of the unweighted and weighted study of the \MLV Power Grid: a general increase in the number of nodes traversed by following the shortest paths between the unweighted and weighted situation. However, the most interesting part of the study deals with an integration of all the parameters investigated with the \CNA to provide a measure that associates the topological quantities to the economical aspects that might influence the spread of a distributed energy exchange market. We combine the topological measures to have two quantities: one represents the losses experienced in the network (a function of the weighted paths and nodes traveled) and the other represents the reliability and redundancy of the network (a function of the disruption behavior and of the redundancy of available paths). With these indicators we show which network samples are more appealing for a distributed energy exchange market.

In~\cite{xiao11} a mainly modeling study is conducted on the cascading failure behavior of networks whose connectivity properties are modeled using the principles of percolation theory. The model used does not take into account real or synthetic Grids to apply the findings, while it bases on the assumption that the \PG can be modeled by a random geometric graph. The aim of the work is to give the analytical conditions (unlike other works that focus on the empirical application on real or synthetic \PG networks) concerning the existence and non-existence of the largest connected component in a network when subject to link failures which are node degree dependent. The author proofs under which conditions a cascading failure exists and when links can be considered \textit{vulnerable}, \textit{reliable} and \textit{unreliable}.

The study of Sun~\cite{sun05} has more the flavor of a survey and a state of the art description than a new analysis.

Pahwa \etal~\cite{pahwa10}, in addition to the vulnerability propose three mitigation strategies in order to improve the network behavior against cascading effects. 1) a load reduction (by a given percentage) in the load supported by each node in the network after the initial failure, so that nodes and links operating below maximal capacity can deal with the redirect load; 2) an idea similar to 1) in the sense of load reduction, but this is applied only to the nodes along the tree that has his root in the node connected to the faulted line and reaches the leaves in the path; 3) considers the situation of isolating a cluster of the network in case of a starting cascade event with the assumption that the cluster can be self-sufficient in terms of energy production (e.g., by using renewable sources). The results shown are extremely encouraging in both the number of nodes that remain in the network (that are not involved in the cascade) when the load is preemptively slightly reduced.

For the sake of completeness, we also mention studies about CNA which have a minor focus on the Power Grid. Most often, the \PG is used as a possible example. In particular, Amaral \etal~in~\cite{Amaral2000} show a study of the Southern California \PG and the model follows an exponential decay for node degree distribution. Watts and Strogatz in~\cite{Watts98} show the Small-world phenomenon applied to the Western States Power Grid while Newman, within a more general work~\cite{Newman2003}, shows the exponential node degree distribution for the same Grid as an example. On the other hand, Barabasi \textit{et al.}~\cite{Barabasi1999a} model the \PG as a scale-free network characterized by a \Pl node degree distribution.

\section{Discussion and Conclusion}\label{sec:discConc}

The survey of CNA studies and their comparison shows how
important properties of a real system such as the \PG can be studied
using graph modeling tools and which conclusions about the
reliability of the infrastructure can be drawn. \CNA proofs to be an
excellent set of tools that provide, although without dealing with the
details and complexities of the electrical properties in the case of the Power Grid, a
comprehensive and general understanding of the properties that characterize
a network.
We see an interesting trend in the various works analyzed, that is, the research towards more complex representation of the properties of the network than a simple graph. In fact, although \CNA can help in understanding the foundational properties of the network of the \PG it is always worth to remember that the \PG is subject to the law of physics and the principles of electrotechnics. From the initial studies (e.g.,~\cite{Watts03,alb:str04,Chassin05,holmgren06,sun05}) considering the \PG just as an indirected  graph without any property (i.e., weight) on edges and with no characterization of the nodes, more recent studies take into account the electrical properties of the \PG system. Of course the aim of these later studies is always to provide anyway a simplification of the highly complex Power System, but they add those essential parameter to better simulate the \G behavior: impedence parameters associated to the transmission lines, power limits supported by the substations (i.e., nodes) such as for instance in~\cite{Mei11,bompard09,hines10,guohua08,wang10_c}. The inclusion of a more detailed description of the \PG under investigation enables to better understand the dynamics guiding the \PG with a mixed approach: both preserving the idea of the \CNA of having a general and statistical behavior of the overall behavior of the Power Grid, and, on the other hand, to take into account the physical/electrical properties essential to characterize the Power Grid. Latest results in~\cite{hines10,bompard09,arianos09} show a better agreement to real \PG behavior of the models that take into account physical parameters, compared to the pure topological analysis, and the observed behaviors and critical points in real power systems thus justifying this \textit{enhanced \CNA} approach.

A noteworthy general aspect is the role that \CNA has in the \PG infrastructure vulnerability analysis: CNA does not want to substitute the traditional approaches to Power Systems resilience and safety analysis since they have proved extremely successful in governing and managing the it with only occasional highly disruptive events. We stress once again that \CNA techniques applied to the \PG world are high simplification of all the complexities governing the Power Systems, but it can anyway be helpful to give a general vision that can help in identifying quickly and in a simple way critical spots or aspects of the \PG which then may be further and deeply analyzed with traditional electrical engineering tools.

A recent emerging function that CNA is acquiring deals with the design aspects of the Power Grid~\cite{pag:preprint12}. In fact, CNA can be used in simulating new \G topologies, assessing the strategies to evolve actual topologies to improve them, especially in the panorama of the forthcoming Smart Grid that might require a reshape of the Grid especially in the \MLV part.

%
%

The samples almost entirely belong to the \HV network and are part either to the American \G (or subsamples of it) or to the European one (or part of it) or Chinese Grid. Generally, the node degree distribution tends to follow an exponential distribution with some minor exceptions, allowing a general characterization of the properties of the network by the average degree parameter. Betweenness studies are also interesting to characterize the criticality of certain nodes as essential for the ability of the network to guarantee its navigability. Unfortunately, the computation of betweenness is done only by a small part of the studies under assessment. However, the tendency is to have at least for \HV samples a behavior that is closer to a \Pl distribution for the probability characterizing this metric. It means having very few nodes with very high values that are responsible to allow the majority of the shortest paths across the network. On the other hand the results from the different studies contrast regarding the \sw phenomenon in \PG networks. Indeed, some of the conditions imposed by Watts and Strogatz~\cite{Watts98} are not met by \PG samples due to physical and economic reasons. This property must depend on the specific sample analyzed.
The geography of the country whose sample is derived is sometimes important (e.g., Italy~\cite{Rosato2007}) while for other studies it has lower impact~\cite{Corominas-murtra2008}. A point of agreement between all the studies is about the reliability of the \PG networks when facing failures. A general good resilience to random breakdown, while extreme vulnerability is experienced by attacks that target the critical nodes (i.e., high node degree or high betweenness nodes). 

An important result for the accuracy of the \CNA studies is the similarity of results that this type of analyses give compared to the traditional electrical engineering results~\cite{Chassin05,Corominas-murtra2008,arianos09,bompard09}. This really shows how the theoretical study and the measured quantities in the real environment are very close.

Of course to enable \CNA to be a more useful tool in the Power Systems domain, especially in order to create models that take into account the physical parameters of the network data and information related to the electrical parameters, it is essential for CNA and \PG scientists to have precise and reliable information of the topology and characteristics of the electrical Grid. As remarked by Rosas-Casals~\cite{casals10} in order to better model and more precisely explain the behavior of the Power Grid, high quality and precise information are required about the Grids, their components, the actual flow of energy data the \Gs deal with in their operations, the properties of the power lines and also information for those part of the Grid which are held privately in the hands of Grid operators and distribution operators. Security and criticality of this data is a recognized aspect in a world that fears terrorist attacks, but without data feeding the models and research works it is not possible to discover actual vulnerability points and help to build more secure and reliable Power Grids.

Networks are an integral model of phenomena surrounding us, may these
be biological networks (e.g.,
food-webs, protein interactions) or human generated ones  (e.g., airline
travel routes, computer chip wiring, telephone call
graphs)~\cite{bar:lin03}. Having methods and tools to better
understand them and their dynamics 
is beneficial for knowledge advancement and better design of future
systems. \CNA is such a modeling technique that provides
methods and metrics for an analytical comprehension of network
behaviors. Public infrastructures are important for today's society,
in particular the Power Grid, which is by nature a complex network,
has a critical role for the economy in every country. Having an
overall view of the \PG as a Complex System gives the ability to
assess the potential issues the electrical system may face  due to
topological failures. In this paper, we have shown what are the main
studies conducted on different \PG networks using metrics and
techniques from the emerging CNA field of study. Although the basic
methodology of study is the same, indeed the \CNA techniques, the
results show differences in some properties such as node degree
distribution statistics, the presence or absence of the \sw
property. On the other hand, the commonality is the behavior of the
\PG networks when facing failures: a general good resilience to random
breakdown, while they show extreme vulnerability when facing attacks
that target the critical nodes (i.e., high degree or high betweenness
nodes, nodes or lines that manage the highest electricity flow). The morphology of the country definitely influences the the
topology of the network and thus its properties and reliability,
symbolic is the case of the Italian Grid~\cite{Rosato2007}.

To have a more complete idea of the \PG networks it is worth
investigating other Grids from other countries, not only limiting
oneself to networks in Europe and United States and China as the studies
analyzed so far focus on. The investigation of Grids belonging to other geographies such as
Asia and South America could lead to new topologies. On the other hand, it is important to study
more samples belonging to the \MLV Grids as to the best of our
knowledge the only study in this direction is our own
\cite{PaganiAielloTSG2011}. This is interesting not only because it
highlights some different properties from the High Voltage, but also
because it can provide indications useful for the design of the future
Smart Grid. The analysis we performed also takes into account 
the weights representing physical properties of the Grid and is
therefore more informative about transport capabilities of the
distribution network.  In addition, \CNA can be used not only
as a tool for the analysis of the Grid, but also to consider how the
electrical Grid might evolve according to design principles to be
optimized at a topological level~\cite{Wang2010,pag:preprint12}. It is also interesting taking into
account the influence of the network topology on electricity
distribution costs for the future scenarios of Smart Grid solutions~\cite{PaganiAielloTSG2011}.

\section*{Acknowledgment}
The work is supported by the EU FP7 Project
GreenerBuildings, contract no. 258888. The work is also supported by the Dutch National Research
Council under the NWO Smart Energy Systems programme, contract
no.~647.000.004. Pagani is supported by University of Groningen with the Ubbo Emmius
Fellowship 2009.

\bibliographystyle{IEEEtran}
\bibliography{andrea.bib}

\end{document}